\documentclass[11pt]{article}
\usepackage[final]{acl}

\usepackage{times}
\usepackage{latexsym}

\usepackage[T1]{fontenc}

\usepackage{microtype}

\usepackage{inconsolata}

\usepackage{graphicx}

\usepackage{graphicx}
\usepackage{wrapfig}
\usepackage{booktabs}
\usepackage[table]{xcolor}
\usepackage{colortbl} 
\usepackage{placeins}
\usepackage{subcaption}
\usepackage[textsize=tiny]{todonotes}
\usepackage{pifont}
\usepackage{multicol}
\usepackage{array}
\usepackage{wrapfig}
\usepackage{subcaption}
\usepackage{multirow}
\usepackage{enumitem}
\usepackage{listings}
\usepackage{fontawesome5}

\usepackage{mdframed}
\usepackage{tikz}

\usepackage[moderate,leading=normal,tracking=normal,wordspacing=normal,lists=normal]{savetrees}

\definecolor{errorred}{RGB}{211,111,100}
\definecolor{errorline}{RGB}{255,235,238}
\definecolor{codebg}{RGB}{250,250,250}
\definecolor{commentgray}{RGB}{120,120,120}
\definecolor{darkgreen}{rgb}{0.0, 0.5, 0.0}
\definecolor{verylightgray}{rgb}{0.97, 0.97, 0.97}

\lstdefinestyle{asmstyle}{
    basicstyle=\ttfamily\scriptsize,
    numbers=left,
    numberstyle=\tiny\color{gray},
    numbersep=5pt,
    frame=none,
    breaklines=true,
    tabsize=4,
    showstringspaces=false,
    xleftmargin=10pt,
    commentstyle=\color{commentgray},
    escapeinside={(*@}{@*)},
    columns=flexible,
    keepspaces=true,
}

\newmdenv[
  linecolor=darkgreen,
  linewidth=0.5pt,
  topline=true,
  bottomline=true,
  leftline=true,
  rightline=true,
  skipabove=5pt,
  skipbelow=5pt,
  innertopmargin=5pt,
  innerbottommargin=5pt,
  innerleftmargin=5pt,
  innerrightmargin=5pt,
  frametitleaboveskip=5pt,
  frametitlebelowskip=5pt,
frametitlebackgroundcolor=errorred,
  frametitlerulewidth=0.5pt,
  frametitlefont=\sffamily\bfseries\small\color{white},
  nobreak=false,
]{errorboxbase}

\newenvironment{errorbox}[1]{%
  \begin{errorboxbase}[frametitle=#1]%
}{%
  \end{errorboxbase}%
}

\definecolor{darkgreen}{rgb}{0.0, 0.5, 0.0} 
\definecolor{verylightgray}{rgb}{0.97, 0.97, 0.97}
\newcommand{\cmark}{\textcolor{darkgreen}{\scalebox{1}[1.0]{\ding{51}}}}
\newcommand{\xmark}{\textcolor{red}{\ding{55}}}  
\newcommand{\err}[1]{{\small\textcolor{black!65}{#1}}}

\title{CASS: Nvidia to AMD Transpilation with Data, Models, and Benchmark}

\author{
    {\bfseries Ahmed Heakl$^{1}$}\quad
    {\bfseries Gustavo Bertolo Stahl$^{\dagger1}$}\quad
    {\bfseries Sarim Hashmi$^{\dagger1}$}\quad
    {\bfseries Seung Hun Eddie Han$^{1}$} \\
    {\bfseries Mukul Ranjan$^{1}$} \quad
    {\bfseries Arina Kharlamova$^{1}$} \quad
    {\bfseries Salman Khan$^{1}$$^{,2}$}\quad
    {\bfseries Abdulrahman Mahmoud$^{1}$}\thanks{$\dagger$ Equal contribution.}\quad \\
    { $^{1}$MBZUAI}\quad
    { $^{2}$Australian National University}\quad
    \vspace{10pt}\\
    \faGithub~\url{https://github.com/ahmedheakl/CASS}\\
    \faLink~\url{https://huggingface.co/datasets/MBZUAI/cass}\\
}
\begin{document}
\maketitle
\begin{abstract}
Cross-architecture GPU code transpilation is essential for unlocking low-level hardware portability, yet no scalable solution exists. We introduce \texttt{CASS}, the first dataset and model suite for source- and assembly-level GPU translation (CUDA~$\leftrightarrow$~HIP, SASS~$\leftrightarrow$~RDNA3). \texttt{CASS} contains 60k verified host-device code pairs, enabling learning-based translation across both ISA and runtime boundaries. We generate each sample using our automated pipeline that scrapes, translates, compiles, and aligns GPU programs across vendor stacks.
Leveraging \texttt{CASS}, we train a suite of domain-specific translation models that achieve 88.2\% accuracy on CUDA$\rightarrow$HIP and 69.1\% on SASS$\rightarrow$RDNA3, outperforming commercial baselines including GPT-5.1, Claude-4.5, and Hipify by wide margins.
Generated code matches native performance in 95\% of cases, preserving both \emph{runtime and memory behavior}. To support rigorous evaluation, we introduce \texttt{CASS-Bench}, a curated benchmark spanning \emph{18 GPU domains} with ground-truth execution.
All data, models, and evaluation tools will be released as open source to support progress in GPU compiler tooling, binary compatibility, and LLM-guided code translation.

\end{abstract}

\section{Introduction}\label{sec:intro}
Graphics Processing Units (GPUs) are foundational to modern machine learning and scientific computing workloads due to their high-throughput parallelism. Nvidia’s Compute Unified Device Architecture (CUDA)~\citep{nvidia2024cuda} has become the dominant programming model for GPU acceleration, but its tight coupling to proprietary hardware introduces severe vendor lock-in: CUDA code cannot run on non-Nvidia GPUs due to incompatible instruction set architectures (ISAs)~\citep{nvidia2021turing}. As a result, organizations with large CUDA-based codebases face steep engineering costs when migrating to alternative platforms. Meanwhile, AMD GPUs, offering potential favorable performance-per-dollar~\citep{amd2024benchmarks,verge2024rx9070}, are increasingly adopted across both data centers and consumer devices~\citep{ft2024cuda}, creating a growing need to execute legacy CUDA programs on AMD hardware without full rewrites in software~\citep{zluda2024}.

In response, AMD introduced the Heterogeneous-computing Interface for Portability (HIP)~\citep{hip}, a C++ GPU API built into the Radeon Open Compute platforM (ROCm) stack~\citep{amd2024rocm}, designed to mirror CUDA’s functionality while supporting cross-platform development. HIP enables a unified codebase for both Nvidia and AMD GPUs. Tools like HIPIFY~\citep{amd_hipify}, a static translator, assist migration by converting CUDA-specific constructs into their HIP equivalents, streamlining adoption of the ROCm stack. However, HIPIFY only operates at the source level and cannot execute precompiled CUDA binaries. Furthermore, it exhibits a high failure rate when converting CUDA programs, highlighting the need for more reliable and lower-level transpilation approaches~\citep{zahid2024testing}.

Translating GPU assembly across vendors is hindered by divergent ISAs and compilation pipelines. Nvidia employs a proprietary toolchain centered on \texttt{nvcc}, producing PTX and low-level SASS~\citep{nvidia2024cuda}, while AMD uses GCN/RDNA architectures compiled via the open-source ROCm stack using \texttt{hipcc}~\citep{amd2024rocm} (Figure~\ref{fig:gpu-compiler-stack} provides a detailed breakdown of both stacks). Bridging this gap at the assembly level is critical for democratizing the hardware computing landscape, transfer of hardware-specific optimizations across vendors, and enabling automation beyond source-level rewrites, especially for legacy CUDA codebases rich in low-level tuning. Our work introduces the first foundation for Nvidia-to-AMD assembly and source translation, focusing on correctness and alignment. While not optimization-aware yet, it paves the way for future systems that preserve and adapt performance-critical patterns across GPU backends.

To address the lack of cross-architecture GPU translation datasets, we introduce \texttt{CASS} (\textcolor{green!70!black}{C}UDA-\textcolor{green!70!black}{A}MD A\textcolor{green!70!black}{S}sembly and \textcolor{green!70!black}{S}ource Mapping), a large-scale corpus of 60k semantically aligned CUDA-HIP source pairs and their corresponding host (CPU-x86 ISA) and device (GPU) assemblies for Nvidia (SASS) and AMD (RDNA3) platforms. Each sample comprises functionally equivalent low-level code across vendors, verified through successful compilation and execution, enabling instruction-level analysis across execution boundaries. Unlike generic code corpora like The Stack~\citep{the-stack}, which lack GPU-aligned and compilable content, CASS provides complete source and binary representations across both GPU compute stacks. To construct CASS, we developed a fully open-source pipeline that scrapes, synthesizes, translates (via HIPIFY~\citep{amd_hipify}), compiles, and aligns GPU code. We evaluate CASS along two dimensions: (1) instruction coverage, capturing diverse SASS and RDNA3 opcodes; and (2) domain coverage, spanning real-world compute kernels from ML, graphics, and HPC. CASS is the first dataset to enable source- and assembly-level translation research for GPU architectures.

To validate the utility of our dataset, we introduce \texttt{CASS-Bench}, the first benchmark tailored to cross-architecture GPU transpilation. It spans 18 diverse GPU domains with execution-verified source and assembly pairs, providing a standardized testbed for future work in low-level translation and performance-aware code generation. Building on this benchmark, we present the \texttt{CASS} model family, a suite of domain-specific large language models fine-tuned for both source- and assembly-level GPU code translation. These models are trained on our curated corpus and demonstrate significant improvements over SoTA proprietary systems such as GPT-5.1~\citep{gpt4o}, Claude-4.5~\citep{claude37}, and traditional tools like HIPIFY~\citep{amd_hipify}, achieving 88.2\% source and 69.1\% assembly translation accuracy. Our contributions are as follows:


\begin{itemize}[leftmargin=5pt]
\item \textbf{CASS Dataset Pipeline.} We designed a scalable pipeline for scraping, synthesizing, translating, and compiling CUDA/HIP code into aligned host and device assemblies across Nvidia and AMD GPUs.

\item \textbf{CASS Dataset.} We introduce \texttt{CASS}, the first large-scale dataset for GPU transpilation, containing 60k semantically aligned Nvidia $\leftrightarrow$ AMD pairs  at both the source (CUDA $\leftrightarrow$ HIP) and assembly levels (SASS $\leftrightarrow$ RDNA3), covering 18 real-world GPU domains. 

\item \textbf{CASS-Bench.} We contribute the first evaluation benchmark for cross-architecture GPU translation, with 369 curated tasks across 18 domains, including functionally verified outputs and aligned CUDA/HIP source and SASS/RDNA3 assembly.

\item \textbf{CASS Models.} We release domain-specialized \texttt{CASS} LLMs trained for cross-architecture code translation. Our 7B model achieves $88$\% source and $69$\% assembly accuracy, outperforming GPT-5.1 and Claude by $22.2$\% on CASS-Bench. Crucially, 95\% of translated assemblies preserve execution runtime and memory compared to native compilation, confirming semantic and performance fidelity.

\end{itemize}

\section{Related Works}\label{sec:related}

\paragraph{Translating from Nvidia to AMD.} The fragmentation of GPU software ecosystems has driven the need for robust CUDA-to-HIP translation tools. HIPIFY~\citep{hipify-guide} statically converts CUDA source code into HIP source code, enabling ROCm compatibility via direct syntax substitution. Operating at a lower abstraction, CuPBoP-AMD~\citep{cupbop-amd} translates NVVM IR to HIP-compatible LLVM IR using the LLVM toolchain~\citep{llvm,clang}, offering more flexible intermediate-level interoperability. Earlier, GPU Ocelot~\citep{gpuocelot} explored dynamic binary translation, recompiling CUDA to AMD/x86 ISAs at runtime. Although innovative, it was limited by poor scalability and high overhead, making it impractical for modern GPU workloads. All these tools have lacked consistent updates to keep up with CUDA advances, suffer from usability issues, and operate only at the source level. 

More recently, ZLUDA~\citep{zluda2024} introduced a runtime for executing unmodified CUDA binaries on AMD GPUs by intercepting CUDA APIs and translating PTX/SASS into AMD-compatible code via LLVM IR. Operating at the IR rather than hardware-assembly level, it cannot exploit Nvidia's backend optimizations below PTX. By contrast, we target direct assembly-to-assembly translation to transfer hardware-specific optimizations not reflected in intermediate representations.

\vspace{-0.5em}
\paragraph{Assembly-to-Assembly Translation.} Translating assembly across ISAs is challenging due to divergent instruction sets and execution models. Recent work employs LLMs for this task, including GG~\citep{heakl2025guaranteed}, a lightweight x86-to-ARM transpiler, and Guess \& Sketch~\citep{guess-sketch}, which integrates symbolic reasoning for ARMv8-to-RISC-V translation. A key factor in their success is large CPU-centric datasets enabling cross-ISA training. Given the lack of such datasets for GPUs, a primary goal of this work is to enable transpilation across GPU vendors.

\begin{table}[t]
\centering

\resizebox{\linewidth}{!}{%
\begin{tabular}{lcccccc}
\toprule
\textbf{Domain/} & \textbf{ComputeEval} & \textbf{Rodinia} & \textbf{SHOC} & \textbf{Poly} & \textbf{Babel} & \textbf{Ours} \\
\textbf{Characteristics} & \textbf{NVIDIA} & \textbf{Bench} &  & \textbf{Bench} & \textbf{Stream} \\
\midrule
CUDA (src) & \cmark & \cmark & \cmark & \cmark & \cmark & \cellcolor{green!15}\cmark \\
OpenCL (src) & \xmark & \cmark & \cmark & \cmark & \cmark & \cellcolor{green!15}\cmark \\
SASS (asm) & \xmark & \xmark & \xmark & \xmark & \xmark & \cellcolor{green!15}\cmark \\
RDNA3 (asm) & \xmark & \xmark & \xmark & \xmark & \xmark & \cellcolor{green!15}\cmark \\
\bottomrule
\end{tabular}
}
\caption{Comparison of Domain/Characteristics across Different Datasets. src: source, asm: assembly. }
\label{tab:cass_vs_rest}
\vspace{-1.5em}
\end{table}
\paragraph{Datasets for CUDA and HIP.} As shown in Table~\ref{tab:cass_vs_rest}, existing benchmarks in the GPU space generally focus on runtime performance, do not target the assembly level, and lack paired/aligned data across Nvidia/AMD codebases. ComputeEval~\citep{nvidia_compute_eval} includes only CUDA code for hardware evaluation. Rodinia~\citep{rodinia-suite} and SHOC~\citep{shoc} provide heterogeneous benchmarks using CUDA/OpenCL/OpenMP but omit AMD code and assembly. PolyBench~\citep{polybench} evaluates compilers with CUDA/OpenCL kernels, yet lacks assembly-level or AMD support. BabelStream~\citep{babelstream} benchmarks HIP/CUDA/OpenCL memory bandwidth but excludes assembly and domain diversity. Hetero-Mark~\citep{hetereo-bench} targets CPU-GPU workloads where GPU code is minimal. The Stack~\citep{the-stack,the-stackv2} contains nearly 200k CUDA files but no AMD coverage or aligned assembly. In contrast, \texttt{CASS} offers 60k semantically aligned CUDA-HIP source and SASS-RDNA3 assembly pairs across host and device, forming the first dataset for cross-vendor GPU assembly translation.

To the best of our knowledge, no existing dataset provides \textit{paired} source- and assembly-level Nvidia-AMD code, hindering effective training and benchmarking.

\begin{figure*}[t]
  \centering
  \includegraphics[width=1\linewidth]{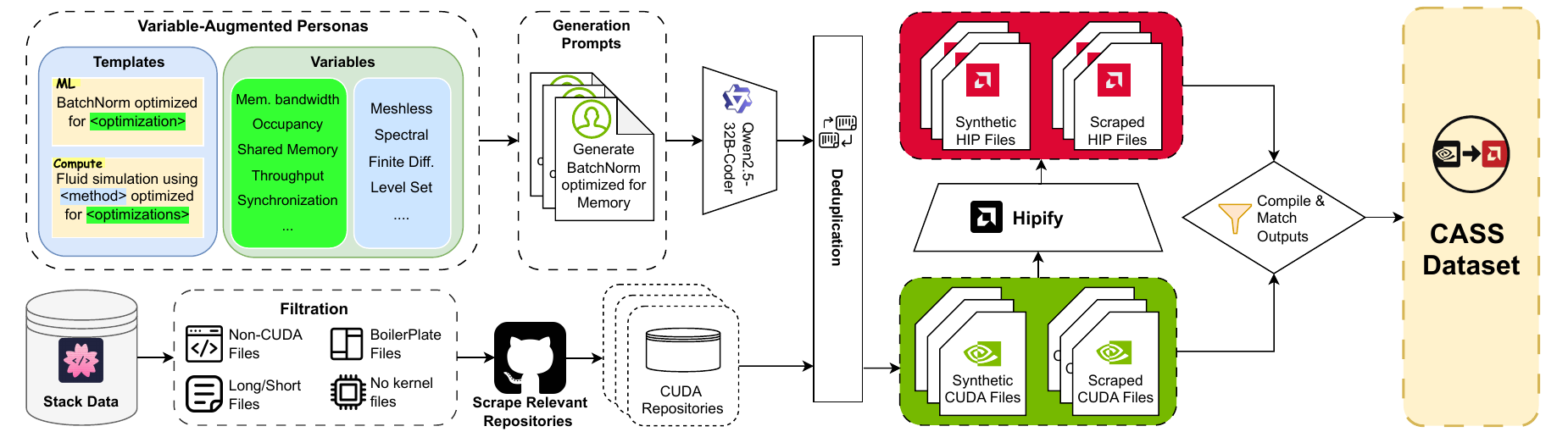}
  \vspace{-0.5em}
  \caption{\textbf{\texttt{CASS} Data Collection Pipeline.} We collect CUDA code from public repositories and synthesize additional samples via prompt-based LLM generation. After filtering and deduplication, all CUDA files are translated to HIP using HIPIFY, then compiled to extract host and device assembly. Matched outputs form the \texttt{CASS} dataset with aligned source and assembly pairs across Nvidia and AMD stacks.}

  \label{fig:cass_pipeline}
  \vspace{-1.0em}
\end{figure*}

\section{Methods}\label{sec:methods}
This section outlines the end-to-end methodology behind \texttt{CASS}, including data collection and code compilation for Nvidia and AMD GPUs. 

\vspace{-0.5em}
\subsection{CUDA Code Scraping}
\label{sec:methods:scraping}

We leveraged the Stackv2 dataset~\citep{the-stackv2} to extract CUDA source files. This dataset, curated from a vast array of public code repositories, offers deduplicated and license-compliant samples, facilitating the assembly of a diverse corpus of GPU-oriented code. To maximize the number of compiled files in the later stage, we used the dataset’s metadata to identify and download the top 200 repositories with the largest number of CUDA files. This repository-level download preserved the original directory structure and relative imports, as shown in Figure~\ref{fig:cass_pipeline}, and improved compilation success by $23.7\%$ compared to isolated file scraping. After extraction, we applied additional filtering to remove overly long files ($>$ 7k lines), trivially short files ($<$10 lines), naive boilerplate samples (e.g., “Hello World”), and files lacking CUDA kernel definitions. This process resulted in a final set of 24k usable samples.

\subsection{Synthetic Data Generation}

To circumvent the issue of low architectural and semantic diversity in underrepresented GPU kernels from ``real-world" code pairs, we employed a coding-oriented large language model (\texttt{Qwen2.5-Coder32B}) to synthesize CUDA kernel code using our variable-augmented persona strategy. We found this important because the amount of GPU code online in general is limited, both in quantity and diversity, and hence one of the goals of \texttt{CASS} is to address this problem for ourselves and others in the space.

The process begins by defining a set of natural language prompt templates with variable placeholders. For example, a template might read: 

\begin{quote}
\textit{Generate a CUDA kernel for cloth simulation with a \{size\}X\{size\} grid. Optimize for \{optimization\}.}
\end{quote}

To fill these templates, we prepared predefined lists of variable values. For instance, \texttt{\{size\}} was instantiated with values such as 32, 64, and 128, while \texttt{\{optimization\}} was sampled from options like ``memory bandwidth", ``register usage", and ``multi-GPU scaling". This allowed us to systematically generate a broad range of prompts, each specifying different values for the placeholders in the templates. Appendix~\ref{suppsec:sythn} includes full details on prompts, variables, and value ranges used for synthetic data generation.

These prompts were then passed to the LLM, which generated CUDA source files accordingly. While this method introduced some functional inconsistencies that required significant post-generation filtering (syntactic errors, missing definitions, or invalid memory operations), it enabled the creation of rich and diverse CUDA samples. In total, we generated 85k CUDA samples, of which 49.1\% compiled successfully, yielding a final set of 20.6k valid files of synthetic data (complementing the 34.1k "real-world" date described in \S\ref{sec:methods:scraping}).

\subsection{Transpilation and Compilation}
\begin{figure}[t]
  \centering
  \includegraphics[width=.95\linewidth]{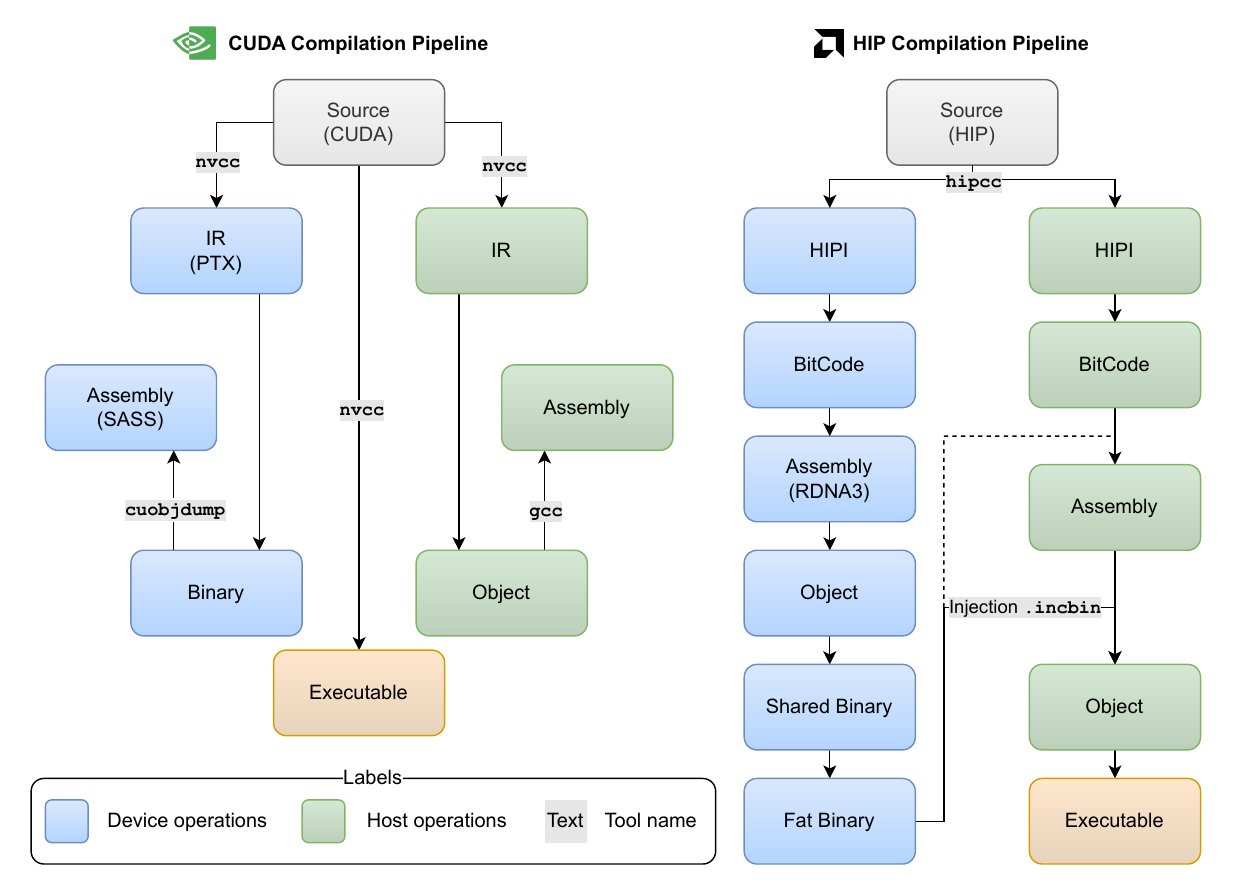}
  \vspace{-0.5em}
  \caption{The Nvidia (left) and AMD (right) stacks illustrate the compilation process for CUDA and HIP. Blue denotes device-side steps; green denotes host-side steps. Nvidia’s stack is opaque; accessing device assembly (SASS) requires first compiling to binary, then using \texttt{cuobjdump}. In contrast, AMD’s process is transparent, allowing direct inspection and modification of device assembly (RDNA3) before host integration.}
  \label{fig:gpu-compiler-stack}
  \vspace{-1.0em}
\end{figure}

After collecting CUDA files from the previous stages, we performed a deduplication sanity check to ensure all samples are unique in our dataset. We then used AMD’s Hipify tool~\citep{amd_hipify} to convert CUDA source files by replacing CUDA-specific API calls with HIP equivalents. Files that failed conversion (approx. 43.9\%) were discarded. Once CUDA-HIP pairs were available, we compiled them to host and device assemblies using \texttt{-Os} compilation flag to reduce code size, achieving a 9.3\% average token reduction compared to \texttt{O3}. Given the architectural divergence of the two stacks (see Figure~\ref{fig:gpu-compiler-stack}), their compilation pipelines differed substantially, requiring significant effort to engineer and standardize our described workflow.

In Figure~\ref{fig:gpu-compiler-stack}, a key distinction between the CUDA and HIP compilation pipelines lies in how they manage host and device assembly separation. In ROCm, the device binary is typically embedded into the host binary during the BitCode-to-assembly transition. We modified this behavior by deferring insertion until after host assembly was converted to object code, enabling: \textbf{(1)} independent extraction of pure host (CPU) and device (GPU) assemblies, and \textbf{(2)} selective recombination for controlled translation and evaluation. 

Conversely, Nvidia provides no access to its binary injection process, device and host assemblies remain intertwined, with no official method for extraction or reintegration~\citep{cuda-binary-utilities}. Since our goal was to support host-to-host and device-to-device transpilation, recombination on the CUDA side was unnecessary. Instead, we developed a regex-based filtering pipeline to disentangle host and device assembly sections during CUDA compilation.

After compiling both stacks to SASS and RDNA3, we retained only samples that compiled successfully on both Nvidia and AMD pipelines, accounting for asymmetric failures. The final dataset includes matched CUDA-HIP source pairs, SASS-RDNA3 device assemblies, and host assemblies. In total, 64k samples were collected after this stage.

\subsection{OpenCL Pipeline}

\begin{wraptable}{r}{0.23\textwidth}
    \centering
    \vspace{-2.7em}
    \setlength{\tabcolsep}{12pt} 
    \renewcommand{\arraystretch}{1.2}
    \caption{Dataset composition by source and size}
    \label{tab:data_composition}
    \resizebox{\linewidth}{!}{
    \begin{tabular}{l@{\hskip 0.5em}r@{\hskip 0.5em}c}
        \toprule
        \textbf{Dataset} & \textbf{Collected} & \textbf{Final} \\
        \midrule
        Synthetic &  85.5k & 20.6k \\
        Stack     & 124.1k & 34.1k \\
        OpenCL    & 6.6k  & 5.9k \\
        \rowcolor[rgb]{0.8,0.94,0.8}
        \textbf{Total} & -- & \textbf{60.7k} \\
        \bottomrule
    \end{tabular}
    }
\end{wraptable}

OpenCL stands as an independent pipeline in generating Nvidia to AMD mapping datasets outside of the CUDA/HIP framework. In other words, it compiles down to the assembly level without going through the aforementioned stacks, operating as a single ``source" for GPU code deveolpment~\citep{KhronosOpenCLGuide}. Approximately 6k OpenCL code snippets were collected from the Stack dataset and compiled down to the device assemblies. On the Nvidia stack, a wrapper C++ function was used to encapsulate the clBuildProgram library provided by OpenCL~\citep{clBuildProgram} and convert them into PTX, after which the CUDA stack was used to compile them down to assemblies. On the AMD stack, clang was used to directly transpile the OpenCL files into device assemblies whilst forcing it to emit intermediate LLVM during this process~\citep{clang}.

\begin{figure*}[t]
  \centering
  \begin{subfigure}[t]{0.40\linewidth}
    \centering
    \includegraphics[width=\linewidth]{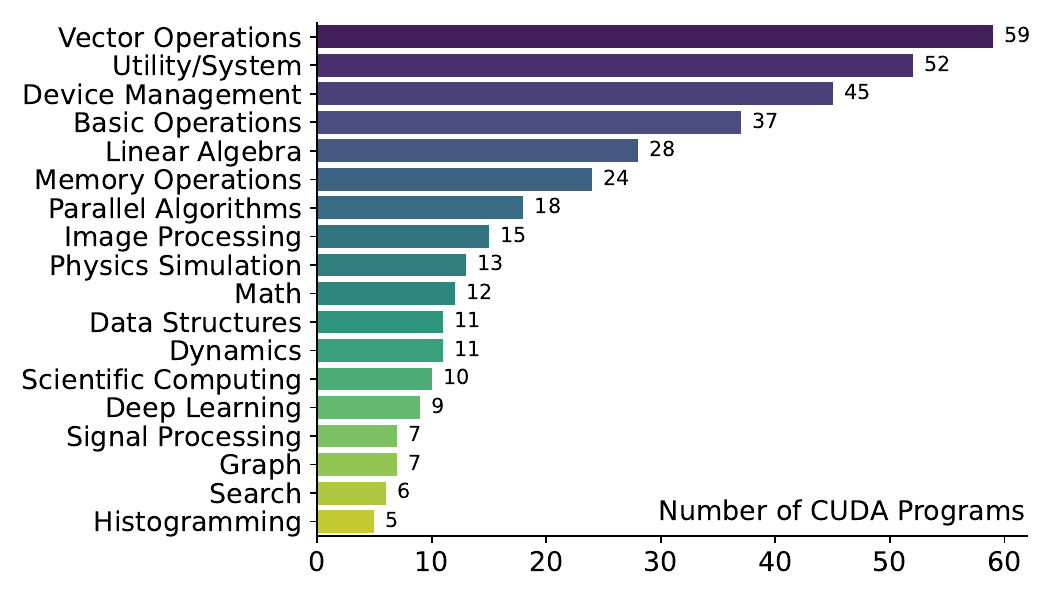}
  \end{subfigure}\hfill
  \begin{subfigure}[t]{0.22\linewidth}
    \centering
    \includegraphics[width=\linewidth,trim=5mm 5mm 5mm 5mm,clip]{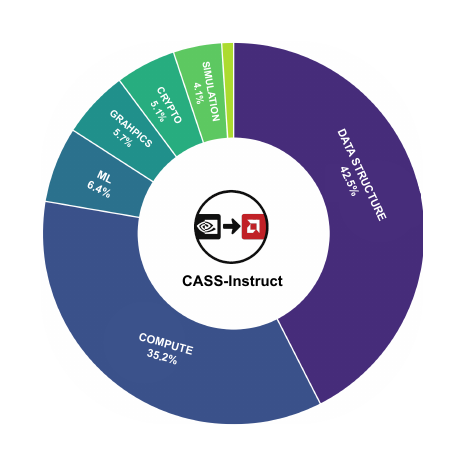}
  \end{subfigure}
  \hfill
  \begin{subfigure}[t]{0.36\linewidth}
    \centering
    \includegraphics[width=\linewidth]{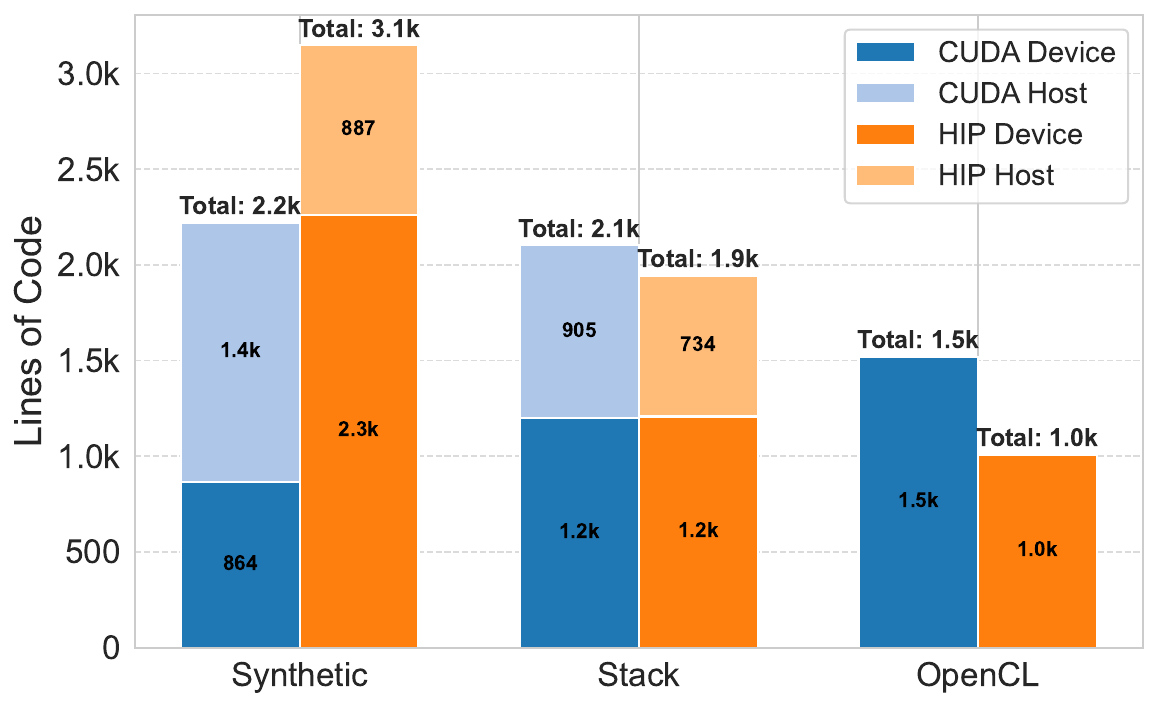}
  \end{subfigure}
  \vspace{-0.5em}
  \caption{\textbf{\texttt{CASS} coverage across dataset and benchmark.} (left) category distribution in \texttt{CASS-Bench} (center) domain distribution of training samples (right) verbosity across subsets and backends.}
  \vspace{-1.0em}
  \label{fig:cass_analysis}
\end{figure*}

The final instruction training dataset (\texttt{CASS}) comprises 60,694 aligned samples spanning a broad range of domains, with a primary focus on GPU compute and GPU-related data structures (Figure~\ref{fig:cass_analysis}, Table~\ref{tab:data_composition}). Each sample includes both CUDA and HIP source code alongside its compiled assembly representation, with pairwise source/assembly alignments verified to compile successfully. All compilations were performed on an Nvidia A100 PCIe machine for the CUDA stack (SASS sm85 ISA) and on AMD Radeon RX 7900 XT GPUs (RDNA3 ISA) for the AMD stack.

\section{CASS and CASS-Bench Datasets}
\label{sec:dataset}

This section presents \texttt{CASS}, a large-scale dataset of aligned CUDA/HIP source and SASS/RDNA3 assembly code, and \texttt{CASS-Bench}, a curated benchmark for cross-architecture GPU translation.

\subsection{Dataset Analysis}

\texttt{CASS} reveals pronounced structural divergence between CUDA and HIP at both source and assembly levels, underscoring the inherent complexity of cross-architecture GPU transpilation. We analyze this by looking at the length of the assembly files, their syntactic similarity, and opcode diversity.

\paragraph{Length of Assembly Files.} Figure~\ref{fig:cass_analysis} (right) shows that AMD device assembly is, on average, twice as long as Nvidia’s in both synthetic and Stack subsets, while Nvidia's device assembly exceeds HIP device assembly by $\sim$50\%  in the OpenCL set. We found an exponential relationship between source complexity and assembly size, with CUDA producing more verbose outputs than HIP for equivalent code. This highlights the growing difficulty of asm-level translation as code complexity scales. See appendix~\ref{suppsec:loc-analysis} for full details.


\paragraph{Opcode Diversity.} Tensor operations dominate both CUDA and HIP assembly, especially in device code, with memory-related instructions such as \verb|mov| and \verb|call| appearing most frequently (refer to appendix~\ref{sec:extra_data_analysis}). HIP opcodes like \verb|s_mov_b32| and \verb|v_add_co_u32| are used extensively, reflecting low-level vector and memory operations unique to AMD's ISA, while Nvidia is dominated by instructions such as \verb|movq|, \verb|call|, and \verb|jmp|, with greater host-side integration. Both stacks share common control and memory ops (e.g., \verb|mov|, \verb|test|), but HIP provides finer-grained access to GPU internals, revealing deeper visibility into parallelism. The synthetic subset emphasizes memory-oriented instructions, aligning with LLM-driven template optimizations. We further conduct a t-SNE visualization of opcode embeddings generated by CodeBERT~\citep{codebert} to examine semantic relationships between Nvidia and AMD instructions. The resulting clusters indicate that, despite backend differences, the two vendors exhibit semantically aligned opcode distributions across device and host levels (Figure~\ref{fig:opcodes_tsne}).

\subsection{CASS-Bench}
\texttt{CASS-Bench} is a 369-sample evaluation benchmark for cross-architecture GPU transpilation, covering both source-level (CUDA $\rightarrow$ HIP) and assembly-level (SASS $\rightarrow$ RDNA3) translation. We constructed the benchmark from the same public repositories used for training data collection (\S\ref{sec:methods}), applying strict deduplication to ensure zero overlap between benchmark and training sets. Each sample was selected to compile and execute successfully on both Nvidia and AMD hardware, with output equivalence verified through automated differential testing. The benchmark spans 18 GPU-centric domains with distribution shown in figure~\ref{fig:cass_analysis} (left).


\section{Experiments}\label{sec:experiments}

We evaluate the \texttt{CASS} dataset by instruction-supervised fine-tuning the \texttt{Qwen2.5-Coder}~\citep{qwencoder} models at various parameter scales. Two variants are developed: one for assembly translation (SASS $\rightarrow$ RDNA3) and another for source translation (CUDA $\rightarrow$ HIP). We benchmark these models against both proprietary and open-source baselines, including larger-scale systems.


\paragraph{Instruction Supervised Finetuning.}\hspace{0.5em} To ensure that input samples fit within the 16K-token context window of the LLM, we normalized CUDA assembly code by removing redundant whitespace and comments, which reduced token count by roughly 15\%. No preprocessing was applied to HIP assembly code due to its sensitivity to whitespace changes. We fine-tuned the \texttt{Qwen2.5-Coder} models at 1.5B, 3B and 7B parameter scales on 4xA100 GPUs, using a batch size of 4, gradient accumulation of 32 (effective batch size of 512) and a learning rate of $1 \times 10^{-5}$. The relatively aggressive learning rate was selected due to the dataset's distributional divergence from the models' pretraining corpus. Training employed DeepSpeed~\citep{deepspeed} with optimizer state sharding to maximize hardware efficiency, achieving 98\% GPU utilization. Additionally, we incorporated Liger Kernel~\citep{liger-kernel} and Paged Adam optimizer~\citep{adamw} to accelerate training and manage memory more effectively. We utilized LLaMA-Factory~\citep{llamafactory} to implement all of these optimizations.  All models were trained with a 16K-token context window. At inference time, we applied RoPE~\citep{rope} extrapolation to support up to 32.7K tokens. Inference was efficient, requiring approximately 56 seconds per a 16K-token sample.

\section{Results}\label{sec:results}

\begin{figure}[!htbp]
  \centering
  \includegraphics[width=0.99\linewidth]{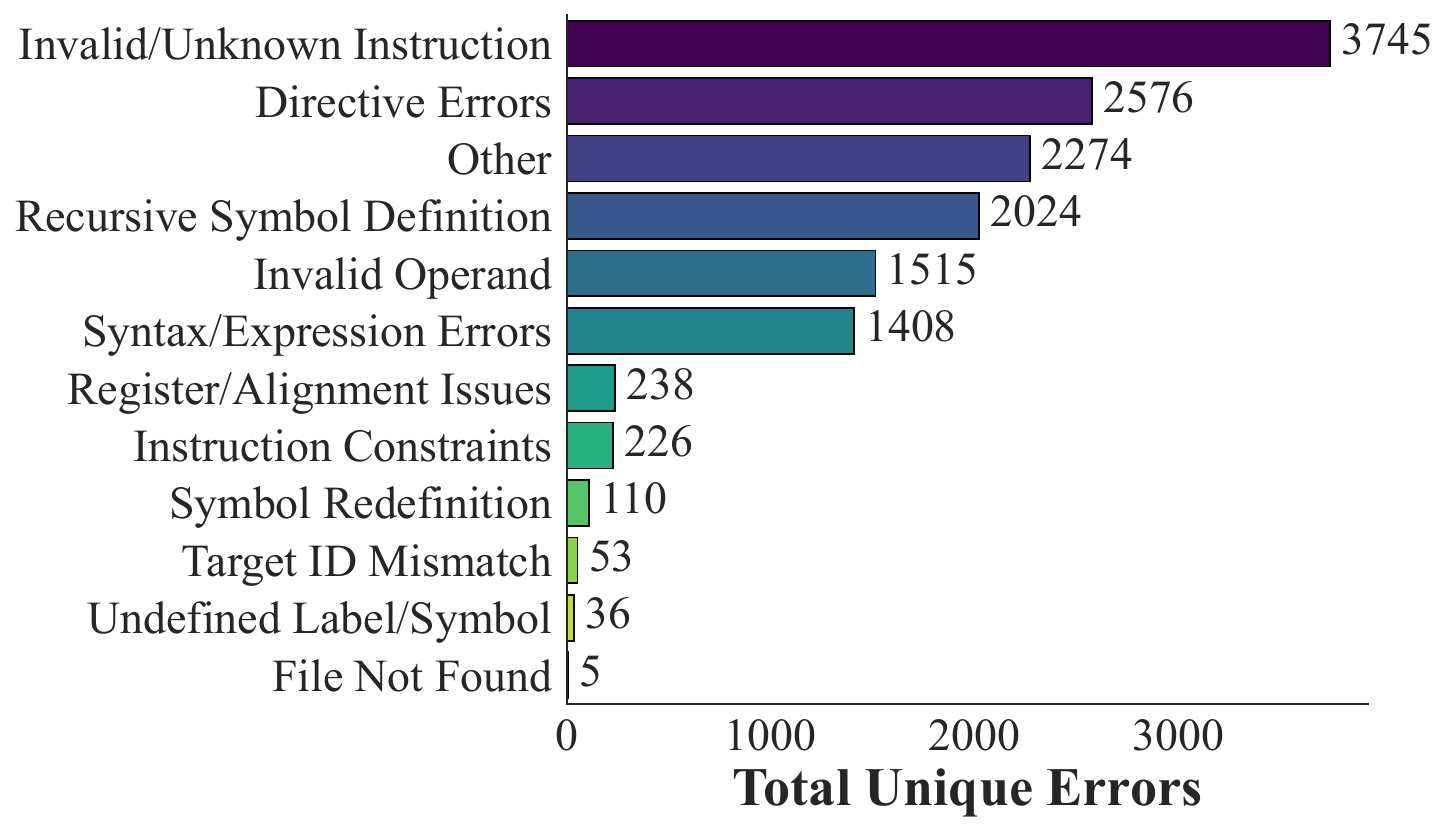}
  \vspace{-0.3in}
  \caption{Distribution of compilation errors.}
  \label{fig:error-distribution}
  
\end{figure}

\begin{table*}[t]
    \centering
    \setlength{\tabcolsep}{4pt}
    \renewcommand{\arraystretch}{0.9}
    \footnotesize      
    \begin{tabular}{@{}llrrrrr@{}}
        \toprule
        & \textbf{Model} 
        & \multicolumn{3}{c}{\textbf{Assembly}} 
        & \multicolumn{2}{c}{\textbf{Source}} \\
        \cmidrule(lr){3-5} \cmidrule(lr){6-7}
        & 
        & \textbf{Test Acc. $\uparrow$} 
        & \textbf{Compile Rate $\uparrow$}
        & \textbf{chrF $\uparrow$} 
        & \textbf{Test Acc. $\uparrow$} 
        & \textbf{Compile Rate $\uparrow$} \\
        \midrule
        
        \multirow{2}{*}{\rotatebox[origin=c]{90}{\textbf{Tools}}} 
        & ZLUDA~\citep{zluda} 
            & 7.86 \err{±0.00} & 8.40 \err{±0.00} & 7.84 & 19.24 \err{±0.00} & 27.59 \err{±0.00} \\
        & Hipify~\citep{amd_hipify} 
            & -- & -- & -- & 87.46 \err{±0.00} & 92.95 \err{±0.00} \\
        \midrule
        
        \multirow{6}{*}{\rotatebox[origin=c]{90}{\textbf{General}}} 
        & GPT-5-mini~\citep{gpt5} 
            & 14.63 \err{±1.68} & 15.72 \err{±3.03} & 14.16 & 66.67 \err{±1.93} & 93.91 \err{±0.98} \\
        & Claude-Haiku-4.5~\citep{claudehaiku45}
            & 11.24 \err{±1.52} & 13.87 \err{±2.90} & 15.31 & 69.42 \err{±1.93} & 95.27 \err{±0.85} \\
        & Gemini-2.5-Flash~\citep{gemini25}
            & 8.94 \err{±1.39} & 10.30 \err{±2.53} & 14.75 & 65.95 \err{±2.02} & 94.98 \err{±0.88} \\
        & LLaMA-3.1-8B~\citep{llama3} 
            & 2.10 \err{±0.67} & 2.89 \err{±1.39} & 8.67 & 28.63 \err{±1.86} & 42.15 \err{±2.02} \\
        & Gemma-3-27B~\citep{gemma3} 
            & 2.44 \err{±0.72} & 2.71 \err{±1.26} & 8.54 & 35.48 \err{±1.96} & 50.18 \err{±2.05} \\
        & Qwen3-30B-A3B~\citep{qwen3} 
            & 16.53 \err{±1.77} & 18.43 \err{±3.16} & 6.79 & 51.25 \err{±2.08} & 78.14 \err{±1.74} \\
        \midrule
        
        \multirow{4}{*}{\rotatebox[origin=c]{90}{\textbf{Reasoning}}} 
        & GPT-5.1~\citep{gpt51} 
            & 22.22 \err{±1.77} & 25.20 \err{±3.66} & 16.62 & 84.95 \err{±1.45} & 98.92 \err{±0.41} \\
        & Gemini-2.5-Pro~\citep{gemini25} 
            & 6.50 \err{±1.14} & 8.94 \err{±2.40} & 14.52 & 66.67 \err{±1.93} & 93.19 \err{±1.04} \\
        & Olmo-3.1-32B~\citep{olmo3}
            & 8.60 \err{±1.35} & 11.11 \err{±2.53} & 9.43 & 27.91 \err{±1.83} & 31.17 \err{±1.89} \\
        & DeepSeek-R1-Qwen-14B~\citep{deepseekr1} 
            & 4.07 \err{±0.93} & 4.34 \err{±1.64} & 5.31 & 8.60 \err{±1.14} & 11.11 \err{±1.29} \\
        \midrule
        
        \multirow{3}{*}{\rotatebox[origin=c]{90}{\textbf{Coding}}} 
        & Qwen3-Coder-30B-A3B~\citep{qwen3} 
            & 1.36 \err{±0.55} & 1.36 \err{±0.88} & 8.21 & 51.25 \err{±2.05} & 78.14 \err{±1.70} \\
        & Qwen2.5-Coder-32B~\citep{qwen25coder} 
            & 5.42 \err{±1.05} & 6.50 \err{±2.02} & 9.65 & 46.95 \err{±2.05} & 73.48 \err{±1.83} \\
        & DeepSeek-Coder-V2-Lite~\citep{deepseekcoderv2} 
            & 2.99 \err{±0.80} & 4.08 \err{±1.64} & 6.21 & 24.17 \err{±1.77} & 35.82 \err{±1.99} \\
        \midrule

        \multirow{5}{*}{\rotatebox[origin=c]{90}{\textbf{Ours}}} 
        & \cellcolor[rgb]{0.95,0.99,0.95}CASS-\texttt{sm89}-1.5B 
            & \cellcolor[rgb]{0.95,0.99,0.95}54.74 \err{±2.05} 
            & \cellcolor[rgb]{0.95,0.99,0.95}67.21 \err{±2.78}
            & \cellcolor[rgb]{0.95,0.99,0.95}81.28 
            & \cellcolor[rgb]{0.95,0.99,0.95}85.91 \err{±1.48}
            & \cellcolor[rgb]{0.95,0.99,0.95}96.75 \err{±0.72} \\
        & \cellcolor[rgb]{0.95,0.99,0.95}CASS-\texttt{sm89}-3B 
            & \cellcolor[rgb]{0.95,0.99,0.95}59.89 \err{±2.02} 
            & \cellcolor[rgb]{0.95,0.99,0.95}81.84 \err{±2.21}
            & \cellcolor[rgb]{0.95,0.99,0.95}79.91 
            & \cellcolor[rgb]{0.95,0.99,0.95}86.99 \err{±1.42} 
            & \cellcolor[rgb]{0.95,0.99,0.95}98.37 \err{±0.51} \\
        & \cellcolor[rgb]{0.9,0.98,0.9}CASS-\texttt{sm80}-1.5B 
            & \cellcolor[rgb]{0.9,0.98,0.9}67.48 \err{±1.89} 
            & \cellcolor[rgb]{0.9,0.98,0.9}74.53 \err{±2.59} 
            & \cellcolor[rgb]{0.9,0.98,0.9}77.53
            & \cellcolor[rgb]{0.9,0.98,0.9}86.74  \err{±1.42}
            & \cellcolor[rgb]{0.9,0.98,0.9}97.49 \err{±0.60} \\
        & \cellcolor[rgb]{0.9,0.98,0.9}CASS-\texttt{sm80}-3B 
            & \cellcolor[rgb]{0.9,0.98,0.9}68.83 \err{±1.93}
            & \cellcolor[rgb]{0.9,0.98,0.9}75.88 \err{±2.40}
            & \cellcolor[rgb]{0.9,0.98,0.9}79.91 
            & \cellcolor[rgb]{0.9,0.98,0.9}87.81 \err{±1.36} 
            & \cellcolor[rgb]{0.9,0.98,0.9}99.28 \err{±0.28} \\
        & \cellcolor[rgb]{0.85,0.96,0.85}CASS-\texttt{sm80}-7B 
            & \cellcolor[rgb]{0.85,0.96,0.85}\textbf{69.11 \err{±1.89}} 
            & \cellcolor[rgb]{0.85,0.96,0.85}\textbf{90.24 \err{±1.70}}
            & \cellcolor[rgb]{0.85,0.96,0.85}\textbf{81.28} 
            & \cellcolor[rgb]{0.85,0.96,0.85}\textbf{88.17 \err{±1.29}} 
            & \cellcolor[rgb]{0.85,0.96,0.85}\textbf{98.92 \err{±0.41}} \\
        \bottomrule
    \end{tabular}
    \caption{
    \textbf{Performance of different models on \texttt{CASS-Bench}.} 
    \emph{Test Acc.} is strict end-to-end accuracy (compile + exact output match) over 369 tasks.
    \emph{Compile Rate} measures number of samples that compiled successfully.
    \emph{chrF} provides character-level n-gram similarity for assembly.
    \texttt{sm80} and \texttt{sm89} denote CUDA compute capabilities for A100 and RTX 4090 GPUs, respectively.
}
    \label{tab:results}
    \vspace{-1.0em}
\end{table*}
\paragraph{Assembly-to-Assembly Performance.} Table~\ref{tab:results} reports \texttt{CASS-Bench} results across LLMs and tools, revealing a striking performance gap. Even frontier models struggle: GPT-5.1 achieves only 22.22\% accuracy, while coding-specialized models like Qwen2.5-Coder-32B reach just 5.42\%. This failure stems from fundamental exposure gaps, SASS is a \textit{proprietary, undocumented} ISA whose opcode-to-binary mappings have only been partially reverse-engineered~\citep{hayes2019decoding}, meaning these models encountered minimal GPU assembly during pretraining. ZLUDA's poor performance (7.86\%) despite operating on compiled binaries reflects its architectural constraints: it translates at the LLVM IR level rather than native assembly, missing backend-specific optimizations like NVIDIA's instruction scheduling and register allocation that exist only in SASS~\citep{zluda2024}. The ISA divergence compounds the challenge, SASS uses fixed 16-byte instructions with implicit stall counts for latency hiding, while RDNA3 employs explicit \texttt{s\_waitcnt} synchronization and separates scalar/vector pipelines~\citep{amd-rdna3-isa}. Our \texttt{CASS-sm80-7B} achieves 69.11\% accuracy by learning these vendor-specific patterns from aligned training pairs, demonstrating that domain-specific fine-tuning can bridge ISA gaps that general-purpose models cannot. Analysis of compilation failures (Figure~\ref{fig:error-distribution}) reveals that invalid instructions and operand constraint violations account for the majority of errors, reflecting the model's difficulty in learning vendor-specific encoding rules. Figure~\ref{fig:error-distribution} further shows that these failures are most prevalent in computation-intensive domains such as vector operations and linear algebra, indicating that complex arithmetic kernels pose the greatest challenge for assembly-level translation. 

\begin{figure}[h]
    \vspace{-0.3in}
    \centering
    \includegraphics[width=\linewidth]{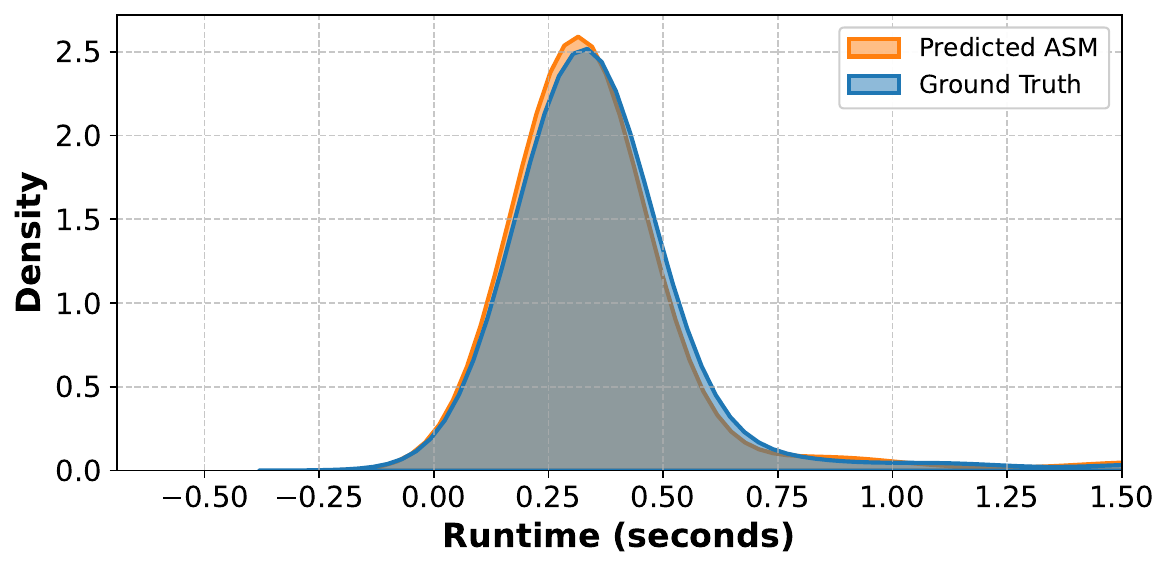}
    \vspace{-2.0em}
    \caption{Runtime distribution comparison between ground truth HIP and CASS-translated assembly. }
    \label{fig:runtime}
    
\end{figure}
\vspace{-1.5em}
\paragraph{Code Efficiency.} 
Figure~\ref{fig:runtime} and Figure~\ref{fig:time-memory-combined} show that the translated code closely matches the original in execution: predicted and ground-truth runtime distributions largely overlap, differing only slightly in the tail. The GPU memory usage deviates by less than $\pm$0.3\,MB, and runtime differences are bounded within $\pm$1.5\,s, with over 95\% of samples falling within $\pm$0.5\,s, confirming that our model preserves both memory and runtime efficiency during assembly translation. Each test was executed 20 times, and the reported values reflect the average across runs to mitigate noise.

\begin{figure}[h]
    \centering
    \includegraphics[width=\linewidth]{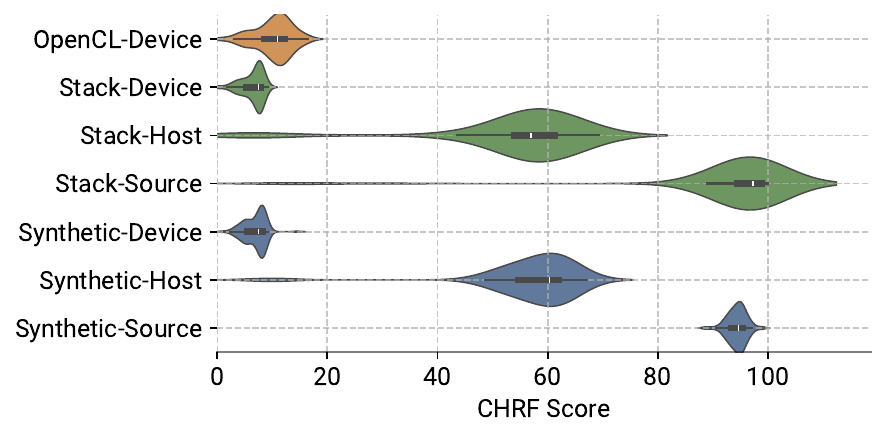}
    \vspace{-1.5em}
    \caption{Syntactic similarity (CHRF) between CUDA and HIP across code levels.}
    \label{fig:chrf_analysis}
    \vspace{-1.3em}
\end{figure}

\paragraph{Source-to-Source Performance.} Source transpilation proves substantially easier than assembly translation, with our best model achieving 88.17\% accuracy compared to 69.11\% for assembly. This gap reflects a deliberate design choice: HIP was engineered as a ``dialect'' of CUDA where most API calls differ only by prefix (\texttt{cuda*}$\rightarrow$\texttt{hip*}), function signatures remain identical, and both languages share the same C++ semantics including templates, lambdas, and classes~\citep{hip-porting-guide}. The high syntactic similarity (CHRF $90\%$, Figure~\ref{fig:chrf_analysis}) means Hipify can rely on regex-based string replacement for the majority of conversions. However, Hipify fails on constructs requiring semantic understanding: inline PTX assembly (which has no HIP equivalent), hardcoded warp sizes (\texttt{32} vs.\ \texttt{64} on AMD), CUDA-specific intrinsics like \texttt{\_\_shfl\_down\_sync}, and architecture-dependent preprocessor guards~\citep{amd_hipify}. Our model surpasses Hipify (88.17\% vs.\ 87.46\%) despite training exclusively on Hipify outputs, a form of ``self-improvement'' where the model learns the underlying semantic mapping rather than mimicking surface-level substitutions, enabling correct translation of edge cases that defeat pattern matching.

\begin{table}[h]
    \resizebox{\linewidth}{!}{
    \begin{tabular}{lccc}
        \toprule
        \textbf{Experiment} & \textbf{Source Acc.} & \textbf{Assembly Acc.} & $\Delta$ \textbf{Impact} \\
        \midrule
        Stack subset                   &  79.67\%  &  48.24\% & - \\
        \rowcolor[rgb]{0.9,0.98,0.9}
        +Synthetic                     &  84.28\%  &  59.89\%  & +11.65\% \\
        \rowcolor[rgb]{0.85,0.96,0.85}
        +OpenCL                       &  87.18\%  &  65.23\%   & +5.34\% \\
        \rowcolor[rgb]{0.8,0.94,0.8}
        +RoPE Extrapolation           &  88.17\%  &  69.11\%   & +3.88\% \\
        \bottomrule
    \end{tabular}
    }
    \caption{\textbf{Ablation study on data sources and context extension.} Each row adds a component cumulatively; $\Delta$ Impact denotes absolute assembly accuracy gain over the previous row.}
    \label{tab:ablations}
    \vspace{-1.3em}
\end{table}

\begin{figure}[h]
    \vspace{-0.5em}
    \centering
    \includegraphics[width=\linewidth]{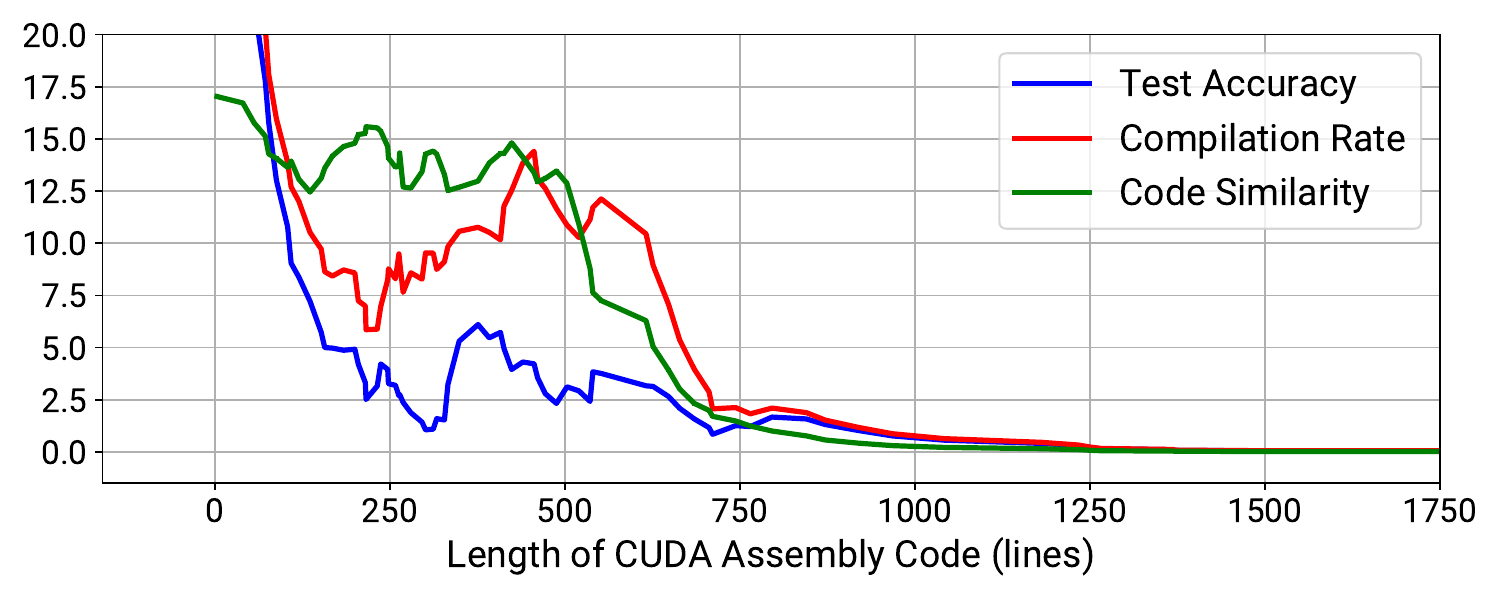}
    \caption{\textbf{Effect of input length on assembly translation performance.} All metrics degrade as sequence length increases.}
    \label{fig:loc-vs-acc}
    \vspace{-1.0em}
\end{figure}

\paragraph{Ablation Study.} Table~\ref{tab:ablations} reveals that synthetic data contributes the largest improvement (+11.65\%), confirming LLM-generated samples provide architectural diversity absent from real-world code (e.g., sparse tensor ops, N-body simulations). OpenCL adds +5.34\% despite only 6k samples, as its distinct compilation pathway introduces different instruction patterns (e.g., explicit \texttt{\_\_local} memory management absent in CUDA). RoPE extrapolation yields +3.88\% by extending context length, beneficial for longer sequences where position encodings degrade (Figure~\ref{fig:loc-vs-acc}).

\paragraph{Hardware Generalization.} To assess cross-microarchitecture transfer, we evaluate our \texttt{CASS-sm80-7B} model (trained on A100 SASS) on RTX4090 (sm89, Ada Lovelace) $\leftrightarrow$ RX7900 XT. Accuracy drops sharply from $69.11\%$ to $32.5\%$, despite both being NVIDIA architectures. This degradation reflects ISA divergence between compute capabilities: \texttt{sm89} introduces FP8 tensor instructions absent in \texttt{sm80}, doubles FP32 throughput per SM (altering instruction scheduling), and employs a 16$\times$ larger L2 cache that changes memory access patterns~\citep{nvidia-ada-tuning}. The compiler also generates different stall counts and register allocation strategies across generations~\citep{jeb-sass-reverse}. That these variations occur \textit{within} NVIDIA's ecosystem yet cause $>$50\% relative accuracy loss underscores a key insight: GPU assembly translation is fundamentally a cross-microarchitecture problem, where even generational changes invalidate learned instruction mappings. This motivates CASS as a living benchmark the community can extend to new ISA pairings as hardware evolves.

\paragraph{Impact of Input Length.} Figure~\ref{fig:loc-vs-acc} reveals a strong inverse relationship between input sequence length and translation quality. As assembly code length increases, both test accuracy and compilation rate degrade sharply, with accuracy approaching zero beyond 750 lines. This trend reflects the compounding difficulty of longer sequences: increased code complexity introduces more intricate control flow and memory patterns, while LLMs inherently suffer from long-context degradation where attention quality diminishes over extended inputs~\citep{liu2024lost}.

\section{Conclusion}
\label{sec:conclusion}
We presented CASS, a large-scale dataset and model suite for cross-architecture GPU transpilation, with 60k aligned source and assembly pairs for Nvidia and AMD. CASS bridges source-to-source (CUDA to HIP) and assembly-to-assembly (SASS to RDNA3) mappings, addressing a key gap in low-level portability. We train the CASS model family, achieving $88.2\%$ accuracy in source translation and $69.1\%$ in assembly translation, outperforming proprietary and open-source baselines. Our transpiled code preserves functional behavior: over $95\%$ of samples match native execution in memory usage and runtime. We introduce CASS-Bench, an evaluation suite spanning 18 GPU-centric domains. All models and data are released open-source to support future work in compiler tooling and hardware interoperability.

\section{Limitations and Future Work}\label{sec:limitations}
Despite this progress, CASS does not yet fully capture certain advanced GPU instruction classes, most notably Tensor Core and matrix-oriented operations. This limitation primarily stems from incomplete support in HIPIFY, AMD’s official CUDA-to-HIP source translator, which currently lacks mappings for WMMA, cuBLAS TensorOp functions, and other low-level CUDA primitives. As a result, kernels relying on Tensor Core instructions or warp-level matrix operations fail to translate or compile successfully. While we explored extending coverage by incorporating additional HIP libraries (e.g., hipBLAS, hipTensor), these efforts yielded only marginal gains due to compilation pipeline constraints. We view this as an opportunity for future expansion. As HIPIFY and the ROCm ecosystem evolve to support a broader range of CUDA functionality, CASS can be extended to include richer tensor and matrix workloads, enabling deeper study of performance-preserving translation across increasingly heterogeneous GPU backends.

\section{Use of Language Models}
Large language models were used in a limited capacity to assist with minor editing and polishing of the manuscript, such as improving clarity and grammar. All technical content, experimental design, results, and conclusions were produced, verified, and finalized by the authors.

\bibliography{custom}

\appendix
\newpage
\section{Appendix}\label{suppsec:appendix}

\subsection{Computational Resources and Environmental Impact}

\paragraph{Hardware used.} All experiments were conducted on two distinct machines to generate architecture-specific outputs. For AMD-related compilation and execution, we used a workstation equipped with an Intel i7-14700KF CPU and an AMD Radeon RX 7900 XT GPU. For Nvidia-related outputs, we used a server with an AMD EPYC 9654 CPU and an Nvidia A100 (80GB) GPU. Furthermore, to ensure consistency and reproducibility across platforms, all file generation was performed within Docker containers tailored to each architecture.

\paragraph{Carbon footprint.} Energy consumption was measured using the \verb|CodeCarbon| tool during fine-tuning experiments conducted on 4$\times$A100 (40GB) GPUs for 2 epochs across 60k samples. The results are as follows:

\begin{table}[h]
\centering
\resizebox{0.8\linewidth}{!}{
\begin{tabular}{lcc}
\toprule
\textbf{Model} & \textbf{Energy (kWh)} & \textbf{CO\textsubscript{2} (kg)} \\
\midrule
CASS-1.5B & 20.13 & 12.11 \\
CASS-3B & 34.27 & 20.72 \\
\bottomrule
\end{tabular}
}
\end{table}

These measurements align with expected energy usage benchmarks for models of similar size and training duration.

\begin{figure}[htbp]
    \centering
    \includegraphics[width=\linewidth, trim=5 5 5 5, clip]{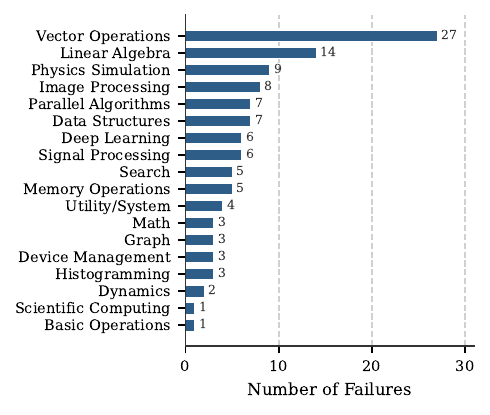}
    \caption{Assembly-level failures across categories.}
    \label{fig:accuracy-breakdown}
\end{figure}

 \begin{figure}[htbp]
    \centering
    \includegraphics[width=\linewidth]{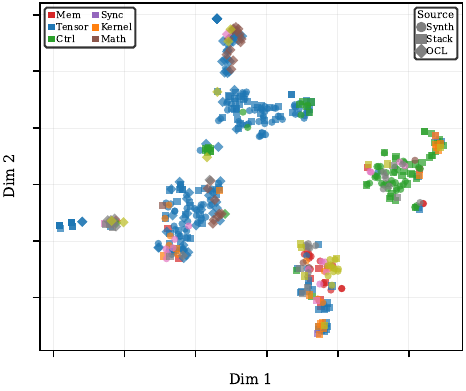}
    \caption{t-SNE projection of CUDA and HIP assembly embeddings.}
    \label{fig:opcodes_tsne}
\end{figure}

\subsection{Prompting Strategy for Closed-source Models}
For the results reported in the paper, we evaluated the assembly-to-assembly translation capacity of closed-source models (GPT-5.1~\citep{gpt51}, Claude-Haiku-4.5~\citep{claudehaiku45}, Gemini-2.5-Flash~\citep{gemini25}, Gemini-2.5-Pro~\citep{gemini25}) and open-source models (LLaMA-3.1-8B~\citep{llama3}, Gemma-3-27B~\citep{gemma3}, Qwen3-30B-A3B~\citep{qwen3}, Olmo-3.1-32B~\citep{olmo3}, DeepSeek-R1-Qwen-14B~\citep{deepseekr1}, Qwen3-Coder-30B-A3B~\citep{qwen3}, Qwen2.5-Coder-32B~\citep{qwen25coder}, DeepSeek-Coder-V2-Lite~\citep{deepseekcoderv2}) using the prompt:

\begin{lstlisting}
You are given a CUDA assembly (SASS) code and 
you are required to convert it into HIP assembly 
(RDNA3) code without changing the functionality. 
The code output from CUDA and HIP should be 
the same when executed. 
Target architecture: {architecture}. 
Return the output in the following format: 
```amdasm
<HIP assembly code>
````. 
CUDA Assembly:
```cudaasm
{cuda_asm}
```
\end{lstlisting}

For the same models, we use the following prompt for evaluating source code: 

\begin{lstlisting}
You are given a CUDA source code and you are 
required to convert it into HIP code without 
changing the functionality. The code output from 
CUDA and HIP should be the same when executed. 
Target architecture: {architecture}. 
Return the output in the following format: 
```amd
<HIP code>
````. 
CUDA Code:
```cuda
{cuda_code}
```
\end{lstlisting}

We use the default inference hyperparameters (\texttt{temperature}, \texttt{top\_p}, \texttt{max\_tokens}) provided by each model. 

We also experimented with more advanced prompting strategies, adding few-shot examples of (SASS, RDNA3) pairs and applying chain-of-thought (CoT) prompting, but observed no significant performance gains. This outcome likely stems from the limited prior exposure these models have low-level GPU assembly code during pretraining. Even with better prompting, the models lack the internal structure or inductive bias needed to reason over hardware-specific instruction patterns.

\subsection{Evaluation on ZLUDA}
To assess ZLUDA’s ability to execute CUDA code on AMD GPUs, we designed a two-track evaluation strategy targeting both source-level and binary-level workflows (the latter being akin to assembly-level translation). In the source-to-source setting, we leveraged access to the original CUDA source files to manually compile them into PTX using \verb|nvcc|. These PTX files were then ingested by ZLUDA, which translated them into AMD-compatible LLVM IR before lowering them into native executables targeting RDNA3 hardware. In the assembly-to-assembly setting, we instead compiled the CUDA source into a complete executable and invoked it directly. ZLUDA intercepted the CUDA runtime calls, dynamically translated the embedded PTX or SASS, and executed the resulting code on the AMD backend. This dual strategy allowed us to assess both ZLUDA’s static translation capabilities and its runtime interoperability under realistic execution conditions.

\begin{figure*}[t]
  \centering
  \begin{subfigure}[t]{0.48\linewidth}
    \centering
    \includegraphics[width=\linewidth]{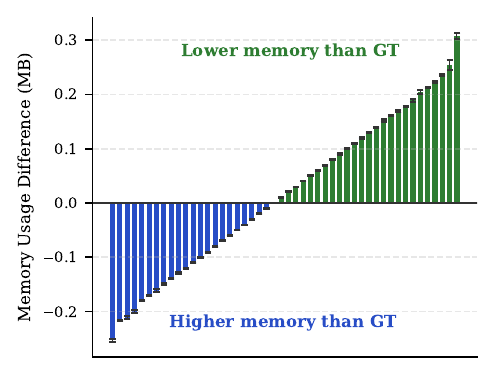}
  
    \label{fig:memory-usage}
  \end{subfigure}
  \hfill
  \begin{subfigure}[t]{0.48\linewidth}
    \centering
    \includegraphics[width=\linewidth]{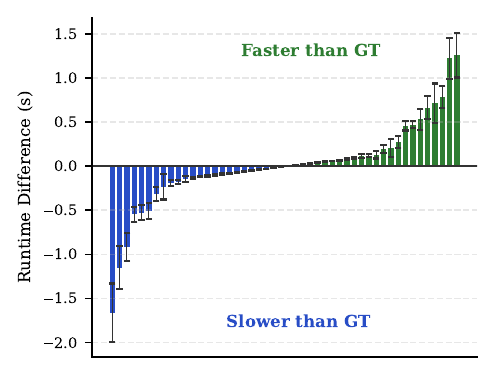}

    \label{fig:execution-time}
  \end{subfigure}
  \caption{Comparison of memory usage (left) and execution time (right) between predicted and ground truth HIP programs, measured via compilation and runtime profiling.}
  \label{fig:time-memory-combined}
\end{figure*}

\begin{figure}[htbp]
  \centering
  \includegraphics[width=0.8\linewidth]{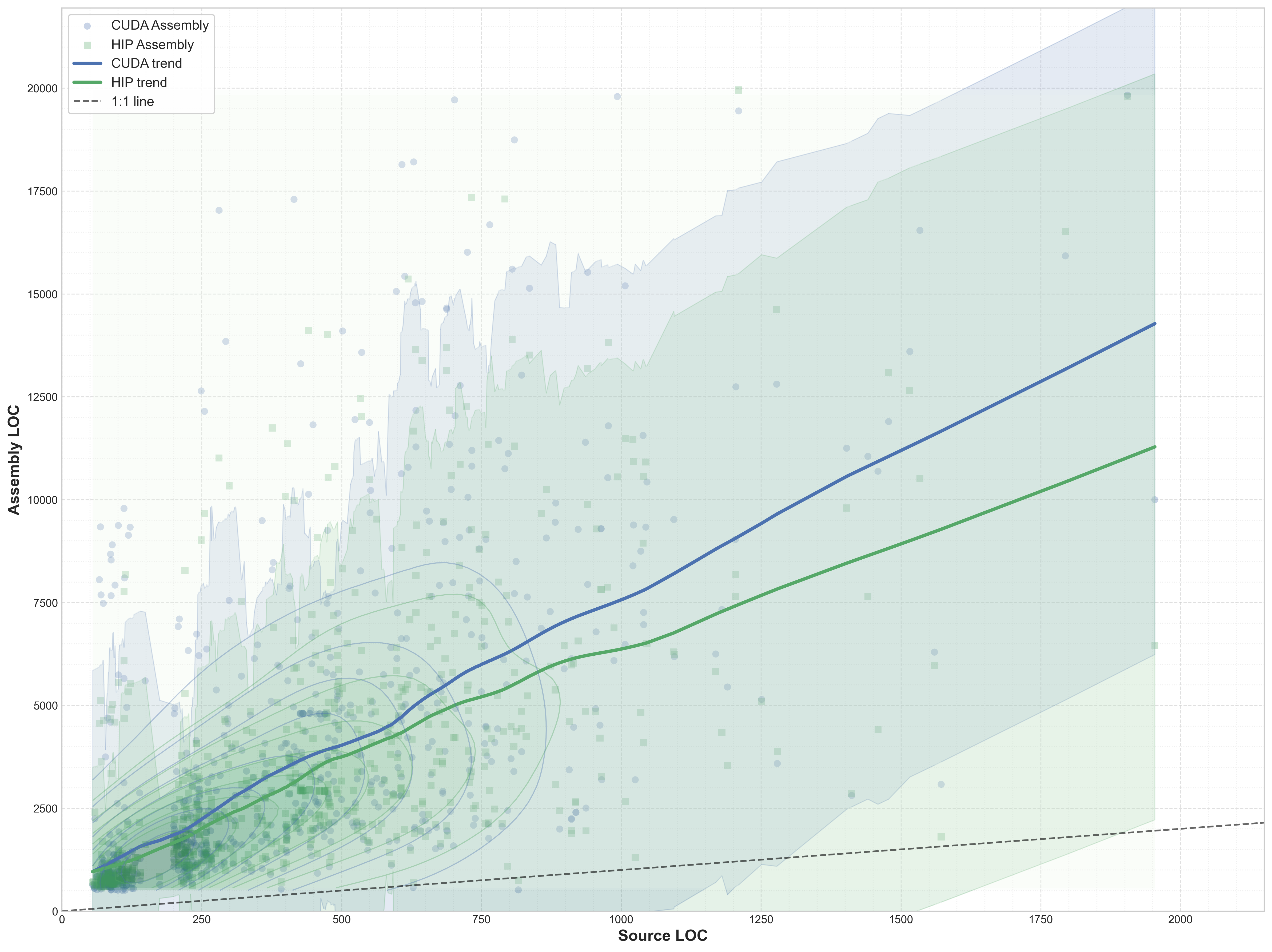}

    \caption{Relationship between source and assembly-level LoC in the CASS dataset. Scatter plot comparing source code lines of code (LoC) to the corresponding assembly LoC for both CUDA and HIP backends across the Stackv2 and Synthetic subsets. Trend lines and density contours illustrate that CUDA typically produces more verbose assembly output than HIP for equivalent source sizes.}  
  \label{fig:loc_compare_asm_vs_source}
\end{figure}

\begin{figure}[htbp]
  \centering
  \includegraphics[width=\linewidth]{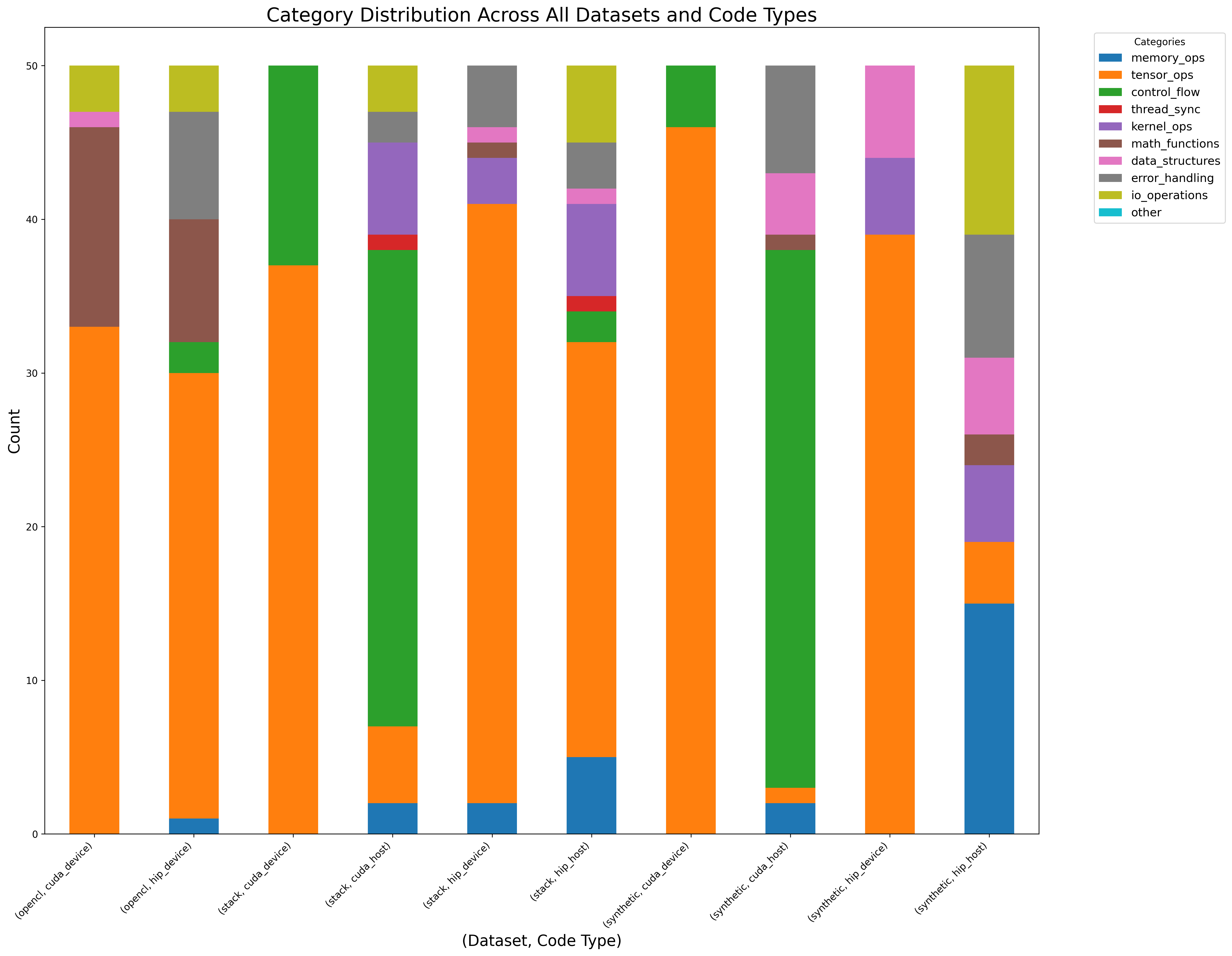}
  \caption{Opcode Category Distribution by Dataset and Code Type. Stacked bar chart showing the distribution of assembly instructions across 10 opcode categories for device and host code in the Synthetic, Stackv2, and OpenCL subsets. Each bar represents a (dataset, code type) pair, illustrating the functional composition of the code across memory, tensor, control flow, synchronization, and other operations.}
  \label{fig:opcodes}
\end{figure}


\begin{figure}
    \centering
    \includegraphics[width=\linewidth]{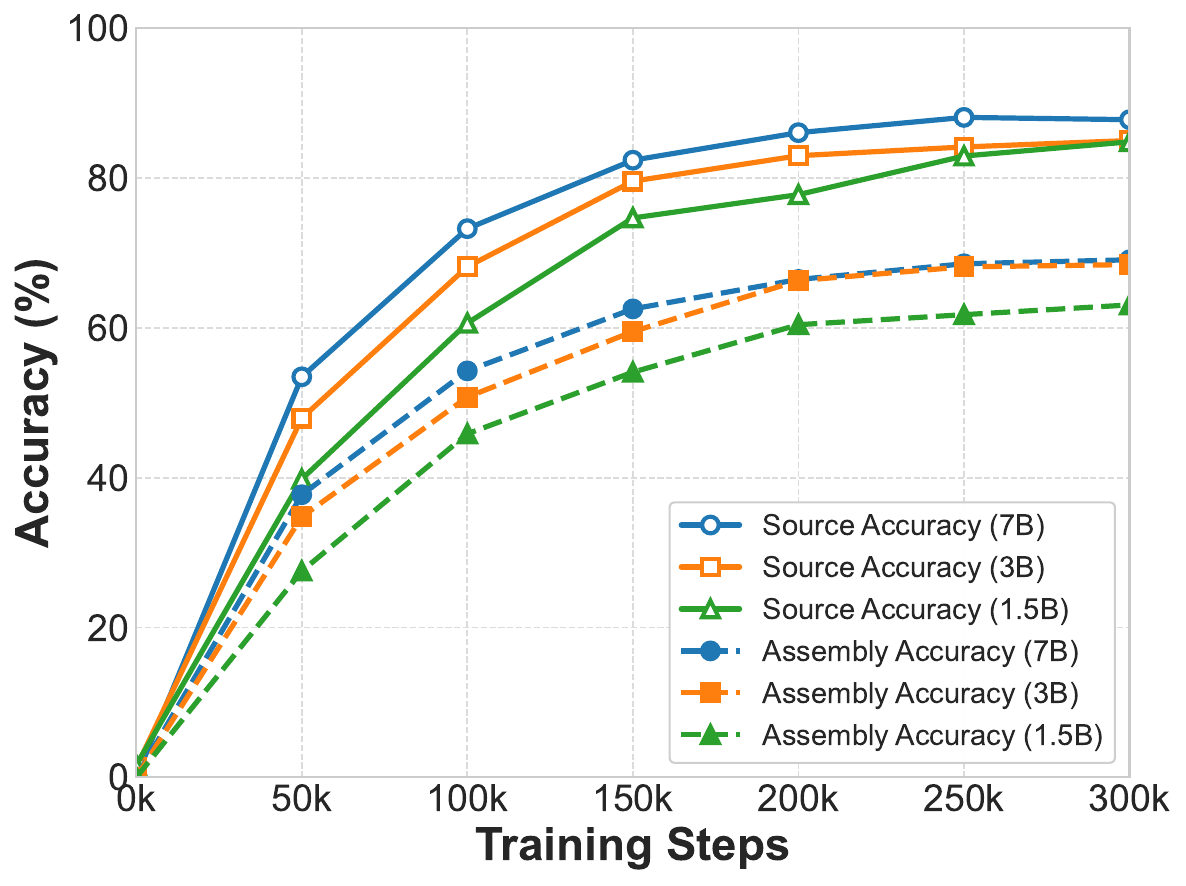}
    \caption{Accuracy vs. training steps for source/assembly across CASS model scales (1.5B, 3B, 7B).}
    \label{fig:data_vs_accuracy}
\end{figure}

\subsection{CASS Domain Coverage}
To obtain the domain-level breakdown shown in Figure \ref{fig:cass_analysis}, we developed a static analysis pipeline that categorizes each source file based on its content. The classification is performed by matching the file’s text against curated sets of domain-specific keywords corresponding to seven high-level categories: \textit{general compute, simulation, data structure, machine learning, graphics, cryptography,} and \textit{scientific computing}. Each keyword set includes terms commonly associated with the respective domain; for example, the \textit{machine learning} category includes terms such as \verb|neural|, \verb|gradient|, and \verb|activation|, while \textit{cryptography} includes \verb|hash|, \verb|encrypt|, and \verb|signature|. For a given file, the domain with the highest keyword match count is assigned. If no keywords are matched, a default label (e.g., \textit{general compute}) is applied. After all files are processed, their assignments are aggregated to produce the final domain distribution. This process provides a simple yet straightforward and interpretable way of grouping source files by their functional domain.


\subsection{Extra Data Analysis}\label{sec:extra_data_analysis}

\subsubsection{Length of Assembly Files}\label{suppsec:loc-analysis}

As shown in the Figure~\ref{fig:loc_compare_asm_vs_source}  We found an exponential relationship between source complexity and assembly size, with CUDA producing more verbose outputs than HIP for equivalent code. This highlights the growing difficulty of assembly-level translation as code complexity scales.

\subsubsection{Opcode Diversity}
In the HIP case, many opcodes, such as \verb|s_mov_b32|, \verb|v_add_co_u32|, and \verb|s_waitcnt|, come directly from AMD’s GPU instruction set. These reflect fine-grained control over the hardware, including scalar and vector operations and synchronization. On the other hand, the CUDA assembly is mostly made up of x86-64 instructions like \verb|movq|, \verb|call|, \verb|jmp|, and \verb|pushq|, which are typically used on the CPU. This suggests that the CUDA output includes more host-side code or that GPU instructions are hidden behind a higher level of abstraction. Still, both stacks share common instructions like \verb|mov| and \verb|test|, showing that some basic control and memory operations are similar. In general, HIP provides more visibility into what the GPU is doing, while CUDA hides many of those low-level details behind a more unified host-device model.

\subsection{Performance Degradation Analysis}
 We analyzed failed kernel translations and categorized them by key operation types. Among the failed cases, we found: 100\% involved control flow (e.g., \verb|for|, \verb|while|), 75\% accessed global memory, 62.07\% used synchronization (e.g., \verb|__syncthreads()|), 10.34\% involved atomic operations (e.g., \verb|atomicAdd|), 6.82\% used shared memory, and 11.36\% included local arrays. It's important to emphasize that these categories are not mutually exclusive, as most failed files involve multiple overlapping operation types. Moreover, as shown in Figure \ref{fig:accuracy-breakdown} in the paper, assembly-level failures are concentrated in Math, Data Structures, Graph, and parts of ML domains, indicating that control-heavy or abstract computation tasks remain the most challenging for the model. These trends suggest that failures are strongly correlated with increased kernel complexity, particularly in terms of global memory access and memory synchronization, which likely strain the model’s limited context and structural understanding at the assembly level.

\subsection{Stratified Benchmarking}
We performed a stratified evaluation of the assembly benchmark by grouping samples based on input length measured in tokens: easy ($<$9k), medium (9k-12k), and hard ($>$12k). The model's accuracy decreases with input length, 35.0\% (easy), 33.33\% (medium), and 17.65\% (hard), highlighting that longer sequences pose greater challenges, likely due to context compression.

\subsection{Sythetic Generation}\label{suppsec:sythn}

To generate large-scale, diverse CUDA programs, we design a multiprocessing Python pipeline that interacts with a locally hosted large language model via a chat-based API. The pipeline leverages a wide array of handcrafted prompt templates, each parameterized with variables such as problem size, optimization target, algorithm type, and architectural features (see Appendix~\ref{sec:prompt-templates}). At runtime, these templates are instantiated with randomly sampled values from curated sets covering domains like matrix operations, graph algorithms, scientific computing, machine learning, and sparse computation (see Table~\ref{tab:placeholder-values}). Each worker process independently generated prompts, sends them to the model, extracts valid CUDA code from the response, and saves the output in a structured format. Robust fault-tolerance mechanisms, including retry logic, output validation, and file existence checks, ensure resilience to model failures and concurrent access. 

Additionally, to avoid reproducing training data seen by the LLM, we apply both prompt-space and output-space deduplication: (1) prompt templates are checked for novelty against LLM training corpora, where verifiable, and (2) generated samples are structurally parsed and filtered using AST and opcode similarity to eliminate near-duplicates. The system supports parallel generation with controlled API concurrency and automatic resumption from previous checkpoints, enabling scalable and efficient generation of compilable CUDA code samples suitable for downstream benchmarking or training.

\begin{table*}[h]
\centering

\rowcolors{2}{white}{gray!10}
\resizebox{\textwidth}{!}{%
\begin{tabular}{ll}
\toprule
\textbf{Placeholder} & \textbf{Example Values} \\
\midrule
\texttt{\{size\}}             & 64, 1024, 16384 \\
\texttt{\{dimension\}}        & 1, 3, 6 \\
\texttt{\{optimization\}}     & memory coalescing, shared memory usage, warp-level programming \\
\texttt{\{operation\}}        & sum, histogram, L2 norm \\
\texttt{\{algorithm\}}        & matrix multiplication, radix sort, BFS \\
\texttt{\{radius\}}           & 1, 5, 13 \\
\texttt{\{graph\_format\}}    & adjacency matrix, CSR, edge list \\
\texttt{\{md\_algorithm\}}    & Verlet integration, leapfrog, Runge-Kutta \\
\texttt{\{linear\_solver\}}   & conjugate gradient, Jacobi, multigrid \\
\texttt{\{numerical\_method\}}& finite difference, spectral, Crank-Nicolson \\
\texttt{\{factorization\_method\}}    & SVD, LU, eigenvalue decomposition \\
\texttt{\{conv\_layer\_count\}}     & 2, 6, 12 \\
\texttt{\{neuron\_count\}}    & 64, 512, 2048 \\
\texttt{\{sparse\_format\}}   & CSR, ELL, HYB \\
\texttt{\{nbody\_algorithm\}} & Barnes-Hut, brute force, particle mesh \\
\texttt{\{filter\_type\}}     & Gaussian, Sobel, Gabor \\
\texttt{\{filter\_size\}}     & 3, 7, 15 \\
\texttt{\{resolution\}}       & 720p, 1080p, 4K \\
\texttt{\{segmentation\_algorithm\}}     & watershed, region growing, U-Net \\
\texttt{\{signal\_transform\}}        & FFT, wavelet, Hilbert \\
\texttt{\{optimization\_algorithm\}}        & Adam, simulated annealing, particle swarm \\
\texttt{\{crypto\_algorithm\}}           & AES, RSA, Argon2 \\
\texttt{\{cracking\_method\}}         & brute force, dictionary attack, rainbow table \\
\texttt{\{hash\_algorithm\}}             & SHA-256, BLAKE3, Bcrypt \\
\texttt{\{data\_structure\}}  & binary tree, hash table, bloom filter \\
\texttt{\{collision\_strategy\}}        & linear probing, cuckoo hashing, separate chaining \\
\bottomrule
\end{tabular}
}
\caption{Representative values for prompt placeholders used in the synthetic code generation.}
\label{tab:placeholder-values}
\end{table*}

\subsubsection{Prompt Templates for Synthetic CUDA Code Generation}
\label{sec:prompt-templates}

The prompts used, listed below, were designed with variations in computation patterns (e.g., memory operations, thread sync, and control flow) and domains (e.g., ML, simulation, graphics), with the intent of diversifying the scope of the synthetic samples.

\subsubsection*{Basic Operations}


\begin{lstlisting}
1. Implement a CUDA kernel for {size}D FFT (Fast Fourier Transform). Optimize for {optimization}.
2. Generate a CUDA implementation for {size}D stencil computation with radius {radius}. Optimize for {optimization}.
3. Write a CUDA kernel for parallel reduction to compute the {operation} of an array of size {size}. Focus on {optimization}.
4. Create a CUDA implementation for convolution operation with a {size}x{size} filter. Focus on {optimization} optimization.
5. Generate a CUDA kernel for matrix multiplication of two matrices A and B of size {size}x{size}. Include error handling and optimize for {optimization}.
\end{lstlisting}

\subsubsection*{Graph Algorithms}


\begin{lstlisting}
1. Write a CUDA implementation for graph coloring of a graph with {size} nodes. Focus on {optimization}.
2. Implement a CUDA kernel for community detection in a graph with {size} nodes using the {community_algorithm} algorithm.
3. Implement a CUDA kernel for graph processing that computes {algorithm} on a graph with {size} nodes. Optimize for {optimization}.
4. Generate a CUDA kernel for finding strongly connected components in a directed graph with {size} nodes. Optimize for {optimization}.
5. Create a CUDA implementation for breadth-first traversal on a graph with {size} nodes stored in {graph_format}. Optimize for {optimization}.
\end{lstlisting}

\subsubsection*{Scientific Computing}


\begin{lstlisting}
1. Write a CUDA implementation for {size}D fluid simulation using {method}. Focus on {optimization}.
2. Create a CUDA kernel for Monte Carlo simulation of {size} paths for option pricing. Focus on {optimization}.
3. Implement a CUDA solver for {size}x{size} sparse linear system using {linear_solver}. Focus on {optimization}.
4. Generate a CUDA implementation for {size}D heat equation solver using {numerical_method}. Optimize for {optimization}.
5. Create a CUDA kernel for molecular dynamics simulation of {size} particles using {md_algorithm}. Optimize for {optimization}.
\end{lstlisting}

\subsubsection*{Machine Learning}


\begin{lstlisting}
1. Generate a CUDA kernel for k-means clustering of {size} data points in {dimension}D space. Optimize for {optimization}.
2. Implement a CUDA kernel for {size}x{size} matrix factorization using {factorization_method}. Optimize for {optimization}.
3. Create a CUDA implementation for computing attention mechanism in a transformer with {size} tokens. Focus on {optimization}.
4. Implement a CUDA kernel for backpropagation in a convolutional neural network with {conv_layer_count} conv layers. Optimize for {optimization}.
5. Write a CUDA implementation for training a neural network with {layer_count} layers and {neuron_count} neurons per layer. Focus on {optimization}.
\end{lstlisting}

\subsubsection*{Sparse Operations}


\begin{lstlisting}
1. Generate a CUDA kernel for sparse FFT computation. Optimize for {optimization}.
2. Implement a CUDA kernel for sparse tensor operations with {size} non-zero elements. Optimize for {optimization}.
3. Write a CUDA implementation for sparse convolution with {size}x{size} filter on sparse input. Focus on {optimization}.
4. Create a CUDA implementation for sparse matrix-matrix multiplication in {sparse_format} format. Focus on {optimization}.
5. Generate a CUDA kernel for sparse matrix-vector multiplication where the matrix has approximately {size} non-zero elements. Optimize for {optimization}.
\end{lstlisting}

\subsubsection*{Simulation}


\begin{lstlisting}
1. Generate a CUDA kernel for cloth simulation with {size}x{size} grid. Optimize for {optimization}.
2. Write a CUDA implementation for raytracing of a scene with {size} objects. Focus on {optimization}.
3. Create a CUDA implementation for {algorithm} of {size} particles in a {dimension}D space. Focus on {optimization}.
4. Create a CUDA implementation for fluid-structure interaction with {size} boundary elements. Focus on {optimization}.
5. Implement a CUDA kernel for N-body simulation of {size} particles using {nbody_algorithm}. Optimize for {optimization}.
\end{lstlisting}

\subsubsection*{Image and Signal Processing}


\begin{lstlisting}
1. Create a CUDA implementation for feature extraction from {size}x{size} images. Focus on {optimization}.
2. Generate a CUDA kernel for image segmentation using {segmentation_algorithm}. Optimize for {optimization}.
3. Write a CUDA implementation for real-time video processing of {resolution} frames. Focus on {optimization}.
4. Implement a CUDA kernel for signal processing with {size}-point {signal_transform}. Optimize for {optimization}.
5. Implement a CUDA kernel for image filtering using {filter_type} filter of size {filter_size}x{filter_size}. Optimize for {optimization}.
\end{lstlisting}

\subsubsection*{Optimization Algorithms}


\begin{lstlisting}
1. Implement a CUDA kernel for simulated annealing with {size} states. Optimize for {optimization}.
2. Generate a CUDA kernel for genetic algorithm with population size {size}. Optimize for {optimization}.
3. Write a CUDA implementation for {optimization_algorithm} with {size} variables. Focus on {optimization}.
4. Write a CUDA implementation for gradient descent optimization with {size} parameters. Focus on {optimization}.
5. Create a CUDA implementation for particle swarm optimization with {size} particles in {dimension}D space. Focus on {optimization}.
\end{lstlisting}

\subsubsection*{Cryptography and Security}

\begin{lstlisting}
1. Generate a CUDA kernel for homomorphic encryption operations. Optimize for {optimization}.
2. Write a CUDA implementation for secure hashing using {hash_algorithm}. Focus on {optimization}.
3. Generate a CUDA kernel for {crypto_algorithm} encryption/decryption. Optimize for {optimization}.
4. Create a CUDA implementation for blockchain mining with difficulty {size}. Focus on {optimization}.
5. Implement a CUDA kernel for password cracking using {cracking_method}. Optimize for {optimization}.
\end{lstlisting}

\subsubsection*{Data Structures}


\begin{lstlisting}
1. Create a CUDA implementation for priority queue with {size} elements. Focus on {optimization}.
2. Create a CUDA implementation for {data_structure} with {size} elements. Focus on {optimization}.
3. Implement a CUDA kernel for operations on a B-tree with {size} nodes. Optimize for {optimization}.
4. Generate a CUDA kernel for skip list operations with {size} elements. Optimize for {optimization}.
5. Write a CUDA implementation for hash table with {size} buckets using {collision_strategy}. Focus on {optimization}.
\end{lstlisting}

\subsubsection{Qualitative Comparison with Other LLMs}

\definecolor{codegreen}{rgb}{0,0.6,0}
\definecolor{codered}{rgb}{0.7,0,0}
\definecolor{codegray}{rgb}{0.5,0.5,0.5}
\definecolor{codepurple}{rgb}{0.58,0,0.82}

\lstdefinestyle{mystyle}{
    commentstyle=\color{codegreen},
    keywordstyle=\color{magenta},
    numberstyle=\tiny\color{codegray},
    stringstyle=\color{codepurple},
    basicstyle=\ttfamily\footnotesize,
    breakatwhitespace=false,         
    breaklines=true,                 
    captionpos=b,                    
    keepspaces=true,                 
    numbers=none,                    
    showspaces=false,                
    showstringspaces=false,
    showtabs=false,                  
    tabsize=2
}
\lstset{style=mystyle}

\definecolor{lightgreen1}{RGB}{189, 232, 201}
\definecolor{lightgreen2}{RGB}{150, 210, 170}
\newcommand{\highgreen}{\makebox[0pt][l]{\color{lightgreen1}\rule[-2pt]{\linewidth}{9pt}}}

\newlength{\highlightwidth}
\newcommand{\highgreenword}[1]{%
  \settowidth{\highlightwidth}{\ttfamily #1}%
  \makebox[0pt][l]{\color{lightgreen2}\rule[-2pt]{\highlightwidth}{9pt}}%
  \texttt{#1}%
}

\definecolor{lightred1}{RGB}{255, 204, 204}   
\definecolor{lightred2}{RGB}{255, 173, 173}   
\newcommand{\highred}{\makebox[0pt][l]{\color{lightred1}\rule[-2pt]{\linewidth}{9pt}}}

\newlength{\highlightredwidth}
\newcommand{\highredword}[1]{%
  \settowidth{\highlightredwidth}{\ttfamily #1}%
  \makebox[0pt][l]{\color{lightred2}\rule[-2pt]{\highlightredwidth}{9pt}}%
  \texttt{#1}%
}

We highlight several cases where \textsc{CASS-7B} outperforms existing LLMs such as Claude, Qwen-Coder, and GPT-4o in faithfully transpiling CUDA to HIP. For example, in one instance, \textsc{CASS-7B} correctly transpiled the CUDA code while preserving the exact string constants from the original program, including the label \verb|CUDA| in the output format string. Maintaining these strings is essential for preserving the intended user-facing behavior, particularly in logging or debugging scenarios where clarity and consistency matter. In contrast, Claude, Qwen-Coder, and GPT4o unnecessarily altered the string to say \verb|HIP|, despite the output still originating from a CUDA kernel. This substitution introduces a semantic error, as the original string refers to CUDA, not HIP, and should remain unchanged.

\textbf{CASS-7B}
\begin{lstlisting}[language=C++,escapechar=\$]
$\highgreen$printf("tanh(%f) = %f $\highgreenword{CUDA}$ vs %f (CPU)\n", 
       h_input[idx], h_output[idx], tanh(h_input[idx]));
\end{lstlisting}

\textbf{Claude, Qwen-Coder, GPT4o}
\begin{lstlisting}[language=C++,escapechar=\$]
$\highred$printf("tanh(%f) = %f $\highredword{(HIP)}$ vs %f (CPU)\n", 
       h_input[idx], h_output[idx], tanh(h_input[idx]));
\end{lstlisting}

In another example, \textsc{CASS-7B} retained the classical CUDA-style kernel launch syntax using triple angle brackets (\verb|<<<...>>>|), while also ensuring that the generated code remained compilable by correctly including the required HIP header \texttt{<hip/hip\_runtime.h>}. This demonstrates a high degree of structural fidelity to the source code, which is especially important for developers familiar with standard CUDA conventions. In contrast, other models such as Claude and Qwen-Coder replaced the launch expression with the HIP-specific macro \texttt{hipLaunchKernelGGL}, which, while functionally valid, deviates from the original representation. More critically, they failed to include the necessary HIP header, rendering the output uncompilable. This example highlights how \textsc{CASS-7B} goes beyond syntactic accuracy to produce code that is both faithful to the original structure and immediately usable in a real compilation setting.

\textbf{CASS-7B}
\begin{lstlisting}[language=C++,escapechar=\$]
$\highgreen$#include <hip/hip_runtime.h>
#include <iostream>
...
$\highgreen$add$\highgreenword{<}$$\highgreenword{<}$$\highgreenword{<}$(N + 255) / 256, 256$\highgreenword{>}$$\highgreenword{>}$$\highgreenword{>}$(d_a, d_b, d_c, N);
\end{lstlisting}

\textbf{Claude, Qwen-Coder}
\begin{lstlisting}[language=C++,escapechar=\$]
$\highred$
#include <iostream>
...
$\highred$$\highredword{hipLaunchKernelGGL(}$add$\highredword{, (}$N + 255) / 256, 256$\highredword{,}$
    $\highredword{0, 0, }$d_a, d_b, d_c, N);
\end{lstlisting}

Lastly, when verifying numerical correctness, \textsc{CASS-7B} preserved the original logging behavior by correctly emitting output to \texttt{std::cout}, as in the source code. This choice maintains consistency with the original program’s semantics, especially in distinguishing between standard output and error streams; important in contexts where output may be redirected or parsed. In contrast, GPT-4o unnecessarily altered the output stream to \texttt{std::cerr}, which, while syntactically valid, changes the runtime behavior of the program. Such a change could lead to unexpected side effects in downstream tools or logging pipelines. This example further demonstrates \textsc{CASS-7B}’s attention to both structural and behavioral fidelity in its translations.

\textbf{CASS-7B}
\begin{lstlisting}[language=C++,escapechar=\$]
$\highgreen$std::$\highgreenword{cout}$ << "Error at element " << i << ": " << h_output[I]
          << " vs. expected " << h_reference[i] << std::endl;
\end{lstlisting}

\textbf{GPT4o}
\begin{lstlisting}[language=C++,escapechar=\$]
$\highred$std::$\highredword{cerr}$ << "Error at element " << i << ": " << h_output[i]
          << " vs expected " << h_reference[i] << std::endl;
\end{lstlisting}

\subsection{Qualitative Analysis of the Failure Cases}
We analyze common failure modes in assembly translation by examining representative cases where the model produces syntactically invalid RDNA3 code. These failures reveal critical challenges in learning GPU assembly constraints that are architectural rather than semantic in nature. The six examples below illustrate distinct categories of errors: register alignment violations (Example 1), bank constraint mismatches (Example 2), instruction hallucination (Example 3), literal operand restrictions (Example 4), control flow inconsistencies (Example 5), and operand type mismatches (Example 6). Each case demonstrates that while the model captures high-level translation patterns, it struggles with vendor-specific encoding rules that are not easily inferred from statistical patterns alone. These constraints, documented sparsely in AMD's ISA manuals and often only enforced at assembly time, represent a key bottleneck in achieving higher accuracy.
\vspace{1.5in}
\begin{errorbox}{Example 1: Invalid Register Alignment}

\textbf{Predicted Assembly:}
\begin{lstlisting}[style=asmstyle]
...
	v_ashrrev_i32_e32 v2, 31, v1
	s_ashr_i32 s5, s4, 31
	s_mov_b32 s3, s4
	s_delay_alu instid0(SALU_CYCLE_1) | instskip(NEXT) | instid1(VALU_DEP_1)
	(*@\colorbox{errorline}{s\_lshl\_b64 s[4:5], s[3:4], 2}@*)
	v_lshlrev_b64 v[0:1], 2, v[1:2]
	s_waitcnt lgkmcnt(0)
...
\end{lstlisting}

\vspace{2pt}
\noindent\textcolor{errorred}{\textbf{Error:}}
\begin{itemize}
    \item \texttt{error: invalid register alignment}
    \item Register pair \texttt{s[3:4]} requires even/odd alignment
\end{itemize}

\vspace{3pt}
\noindent\textbf{Ground Truth (Correct):}
\begin{lstlisting}[style=asmstyle]
...
	s_load_b128 s[4:7], s[0:1], 0x0
	v_ashrrev_i32_e32 v2, 31, v1
	s_delay_alu instid0(VALU_DEP_1) | instskip(SKIP_1) | instid1(VALU_DEP_1)
	v_lshlrev_b64 v[0:1], 2, v[1:2]
	s_waitcnt lgkmcnt(0)
...
\end{lstlisting}
\end{errorbox}

\vspace{8pt}
\begin{errorbox}{Example 2: VGPR Bank Constraint Violation}

\textbf{Predicted Assembly:}
\begin{lstlisting}[style=asmstyle]
...
	v_mov_b32_e32 v4, 0x1b8
.LBB0_2:
	s_delay_alu instid0(VALU_DEP_1) | instskip(SKIP_1) | instid1(SALU_CYCLE_1)
	(*@\colorbox{errorline}{v\_dual\_mov\_b32 v11, v2 :: v\_dual\_mov\_b32 v2, v6}@*)
	s_add_i32 s5, s4, 4
	s_add_i32 s4, s4, 8
...
\end{lstlisting}

\vspace{2pt}
\noindent\textcolor{errorred}{\textbf{Error:}}
\begin{itemize}
    \item \texttt{error: src0 operands must use different VGPR banks}
    \item Dual instructions require operands from different banks
\end{itemize}

\vspace{3pt}
\noindent\textbf{Ground Truth (Correct):}
\begin{lstlisting}[style=asmstyle]
...
	v_mov_b32_e32 v4, 0x1b8
.LBB0_2:
	s_delay_alu instid0(VALU_DEP_1)
	v_mov_b32_e32 v11, v2
	v_mov_b32_e32 v2, v6
	s_add_i32 s5, s4, 4
...
\end{lstlisting}
\end{errorbox}

\vspace{1in}
\begin{errorbox}{Example 3: Hallucinated Instruction}

\textbf{Predicted Assembly:}
\begin{lstlisting}[style=asmstyle]
...
	v_add_nc_u32_e32 v0, 5, v0
	ds_store_b32 v3, v0
	s_waitcnt lgkmcnt(0)
	(*@\colorbox{errorline}{sumba\_send\_msg\_b32 v0, v3}@*)
	s_endpgm
.LBB0_3:
...
\end{lstlisting}

\vspace{2pt}
\noindent\textcolor{errorred}{\textbf{Error:}}
\begin{itemize}
    \item \texttt{error: invalid instruction}
    \item \texttt{sumba\_send\_msg\_b32} does not exist in AMD ISA
\end{itemize}

\vspace{3pt}
\noindent\textbf{Ground Truth (Correct):}
\begin{lstlisting}[style=asmstyle]
...
	v_add_nc_u32_e32 v3, 5, v3
	ds_store_b32 v0, v3
.LBB0_4:
	s_or_b32 exec_lo, exec_lo, s0
	s_waitcnt lgkmcnt(0)
	s_barrier
...
\end{lstlisting}
\end{errorbox}

\vspace{1in}

\begin{errorbox}{Example 4: Multiple Literal Operand Constraint}

\textbf{Predicted Assembly:}
\begin{lstlisting}[style=asmstyle]
...
	s_delay_alu instid0(SALU_CYCLE_1) | instskip(NEXT) | instid1(SALU_CYCLE_1)
	s_or_b32 s12, s8, s2
	s_cmp_eq_u32 s12, s9
	(*@\colorbox{errorline}{s\_cselect\_b32 s12, 0x88, 0x100}@*)
	s_cmp_eq_u32 s13, 0
	s_cselect_b32 s13, s9, 0
...
\end{lstlisting}

\vspace{2pt}
\noindent\textcolor{errorred}{\textbf{Error:}}
\begin{itemize}
    \item \texttt{error: only one unique literal operand is allowed}
    \item Cannot use both \texttt{0x88} and \texttt{0x100} as literals
\end{itemize}

\vspace{3pt}
\noindent\textbf{Ground Truth (Correct):}
\begin{lstlisting}[style=asmstyle]
...
	s_or_b32 s12, s8, s2
	s_cmp_eq_u32 s12, s9
	s_mov_b32 s12, 0x88
	s_cselect_b32 s12, s12, 0x100
	s_cmp_eq_u32 s13, 0
...
\end{lstlisting}
\end{errorbox}

\begin{errorbox}{Example 5: Missing Label Definition}

\textbf{Predicted Assembly:}
\begin{lstlisting}[style=asmstyle]
...
	v_lshl_add_u32 v9, v0, 2, 0
	s_mov_b32 s3, exec_lo
	v_cmpx_gt_u32_e64 v0, v1
	(*@\colorbox{errorline}{s\_cbranch\_execz .LBB0\_8}@*)
; %bb.4:
	v_lshl_add_u32 v2, s4, 3, v9
...
\end{lstlisting}

\vspace{2pt}
\noindent\textcolor{errorred}{\textbf{Error:}}
\begin{itemize}
    \item \texttt{error: undefined label '.LBB0\_8'}
    \item Branch target label was never defined in assembly
\end{itemize}

\vspace{3pt}
\noindent\textbf{Ground Truth (Correct):}
\begin{lstlisting}[style=asmstyle]
...
	v_lshl_add_u32 v9, v0, 2, 0
	s_mov_b32 s3, exec_lo
	v_cmpx_gt_u32_e64 v0, v1
	s_cbranch_execz .LBB0_6
; %bb.4:
	v_lshl_add_u32 v2, s4, 3, v9
...
\end{lstlisting}
\end{errorbox}

\vspace{1in}

\begin{errorbox}{Example 6: Invalid Operand Type}

\textbf{Predicted Assembly:}
\begin{lstlisting}[style=asmstyle]
...
	s_mov_b32 s6, 0
	s_waitcnt lgkmcnt(0)
	s_load_b32 s5, s[2:3], 0x0
	(*@\colorbox{errorline}{v\_cmp\_gt\_u64\_e64 s2, 0x800001000, s[4:5]}@*)
	s_delay_alu instid0(VALU_DEP_1)
	s_and_b32 s2, s2, exec_lo
...
\end{lstlisting}

\vspace{2pt}
\noindent\textcolor{errorred}{\textbf{Error:}}
\begin{itemize}
    \item \texttt{error: invalid operand for instruction}
    \item Immediate \texttt{0x800001000} exceeds 32-bit limit for this encoding
\end{itemize}

\vspace{3pt}
\noindent\textbf{Ground Truth (Correct):}
\begin{lstlisting}[style=asmstyle]
...
	s_mov_b32 s6, 0
	s_waitcnt lgkmcnt(0)
	s_load_b32 s5, s[2:3], 0x0
	v_cmp_gt_u64_e64 s2, s[4:5], 0x7fffffff
	s_delay_alu instid0(VALU_DEP_1)
	s_and_b32 s2, s2, exec_lo
...
\end{lstlisting}
\end{errorbox}
\begin{figure*}[t]
  \centering
  \includegraphics[width=0.95\linewidth]{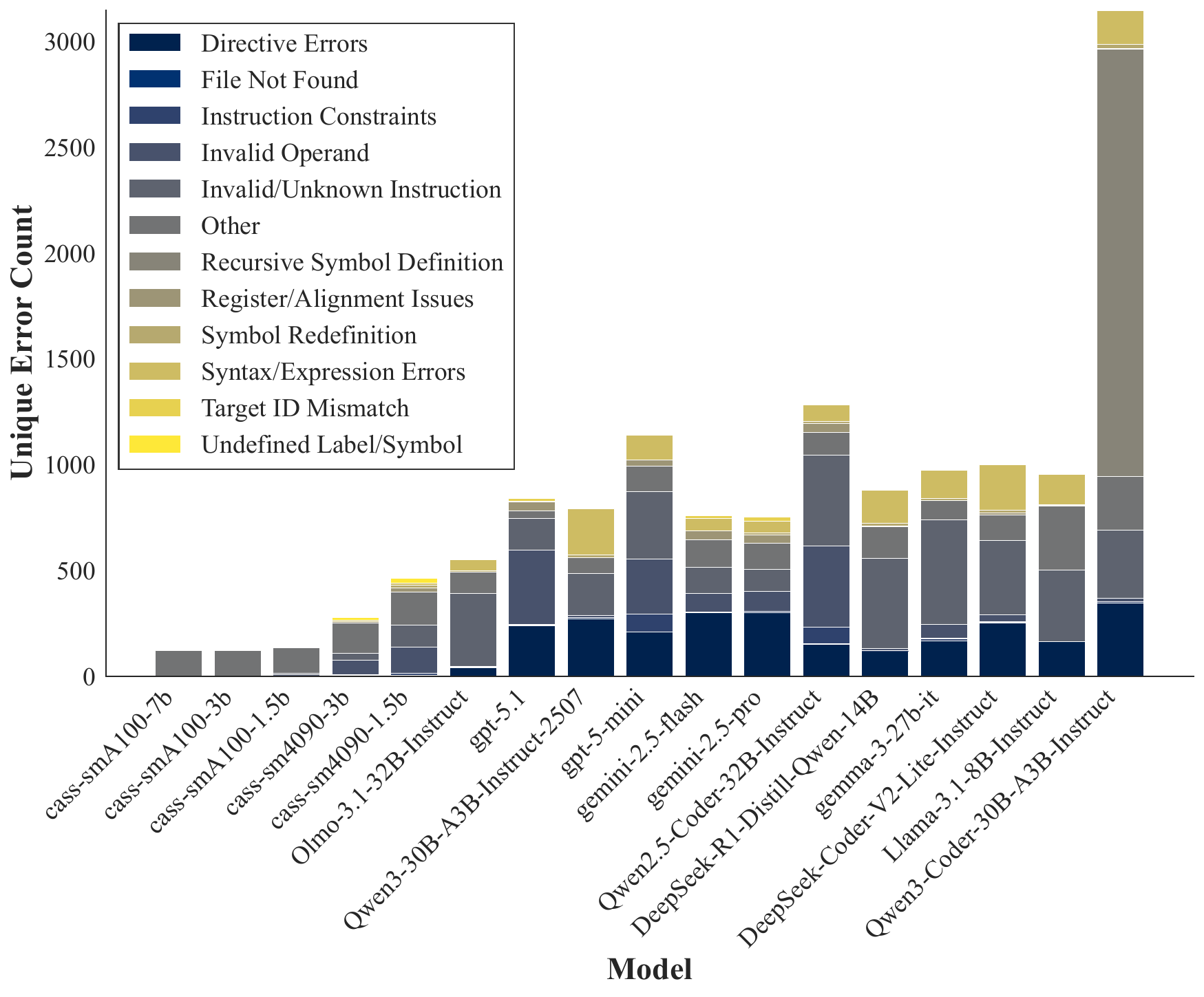}
  \caption{\textbf{Model-wise distribution of compilation errors across different models.} CASS models trained on A100 data exhibit significantly fewer invalid instruction and operand errors compared to all baselines, including frontier models (GPT-5.1, Gemini-2.5-Pro), coding-specialized models (Qwen2.5-Coder-32B, DeepSeek-Coder-V2), and general instruction-tuned models (Llama-3.1, Gemma-3). Scaling from 1.5B to 7B parameters reduces constraint violations by 35-40\%. Models trained on sm89 (4090) show elevated error rates due to ISA divergence.}
  \label{fig:error-distribution-models}
\end{figure*}
\subsection{Quantitative Analysis of the Failure Cases}

To complement the qualitative analysis, we provide a quantitative breakdown of compilation errors across all models and failure cases. Figure~\ref{fig:error-distribution} shows the aggregate distribution of error types, which reveal that invalid or unknown instructions (3745 occurrences) constitute the most common failure mode, followed by directive parsing errors (2576) and operand constraint violations (2024). These patterns confirm that the primary challenge in assembly translation lies not in high-level semantic understanding but in learning the precise encoding rules and instruction constraints specific to RDNA3. Also, register alignment and instruction constraint errors, while less frequent in absolute terms (238 and 226 occurrences respectively), represent categorical failures that prevent any execution, whereas invalid operands may sometimes be caught and corrected through iterative refinement. The relatively low occurrence of control flow errors, such as undefined labels (36 cases), suggests that the model successfully captures basic program structure even when it fails on instruction-level details. This distribution motivates our focus on architectural constraints in the qualitative examples below.

Figure~\ref{fig:error-distribution-models} presents a model-wise breakdown of error distributions across 17 models, enabling direct comparison of failure modes across different architectures, scales, and training paradigms. We observe several patterns from this analysis.  First, our CASS models trained on A100 data (sm80) consistently exhibit fewer invalid instruction errors compared to all baselines, including frontier models (GPT-5.1, Gemini-2.5-Pro, Gemini-2.5-Flash), coding-specialized models (Qwen2.5-Coder-32B, DeepSeek-Coder-V2), and general instruction-tuned models (Llama-3.1-8B, Gemma-3-27B). This suggests that domain-specific training on aligned CUDA-HIP pairs provides better coverage of the RDNA3 instruction space than general pretraining or code-focused curricula. Second, proprietary frontier models show disproportionately high rates of invalid instruction hallucination, with GPT-5.1 and Gemini-2.5-Flash generating over 800 invalid instructions each, nearly 4$\times$ higher than our best CASS model. This is consistent with their limited exposure to GPU assembly during pretraining, as these models were optimized for high-level code generation rather than low-level ISA translation. Interestingly, even coding-specialized open-source models like Qwen2.5-Coder-32B and DeepSeek-Coder-V2-Lite fail at similar rates, indicating that generic code pretraining does not transfer to GPU assembly domains. Third, scaling from 1.5B to 7B parameters within our CASS family reduces invalid operand errors by approximately 40\% and instruction constraint violations by 35\%, indicating that larger models better internalize complex operand encoding constraints and vendor-specific ISA rules. However, the gains diminish beyond 3B parameters, suggesting that architectural knowledge rather than pure capacity becomes the bottleneck.  Fourth, models trained on sm89 (RTX 4090) show elevated rates of instruction constraint violations compared to their sm80 counterparts, with CASS-sm4090-3B exhibiting 28\% more constraint errors than CASS-smA100-3B. This reflects the ISA divergence between Ada Lovelace and Ampere architectures discussed in Section~\ref{sec:results}, where sm89-specific optimizations fail to generalize to RDNA3.  Finally, reasoning-focused models like DeepSeek-R1-Distill-Qwen-14B do not show significant advantages over standard instruction-tuned models, suggesting that assembly translation requires pattern matching over low-level encoding rules rather than high-level reasoning capabilities. The near-uniform performance of general instruction-tuned models (Llama, Gemma, Olmo) at the bottom of the ranking confirms that assembly translation is a specialized skill requiring domain-specific training data.

\end{document}